\documentclass[11pt,a4paper]{article}

\usepackage[a4paper,margin=27mm]{geometry}
\usepackage[T1]{fontenc}
\usepackage[utf8]{inputenc}
\usepackage{amsmath,amssymb}
\usepackage{graphicx}
\usepackage{subcaption}
\usepackage{microtype}
\usepackage{array}
\usepackage{tabularx}
\usepackage{makecell}
\usepackage{float}
\usepackage{pdflscape}
\usepackage[hidelinks]{hyperref}
\usepackage{url}
\usepackage{booktabs}

\title{Large Language Models are Shannon Lossy Compressors Not Solomonoff Induction Estimators:
Self-improvement and Singularity Are Not Near Without Symbolic Model Synthesis}
\author{Hector Zenil$^{1,2}$, Abicumaran Uthamacumaran$^{1}$, Luan Ozelim$^{1}$\\[2mm]
\small $^{1}$ Algocyte | Oxford Immune Algorithmics, Oxford University Innovation \\ \& 
\small London Institute for Healthcare Engineering, UK\\
\small $^{2}$ Algorithmic Dynamics Lab, Department of Biomedical Computing,\\
\small School of Biomedical Engineering and Imaging Sciences, King's Institute for AI,\\
\small King's College London, UK\\
}
\date{}

\newcommand{\E}{\mathbb{E}}

\date{}

\begin{document}
\maketitle

\begin{abstract}
We connect two questions in Algorithmic Information Theory (AIT), Machine Learning (ML) and Artificial General Intelligence (AGI): whether LLMs estimate Solomonoff induction, and whether they can self-improve towards an AI Singularity. We provide theoretical, methodological and empirical answers in the negative but show how limits can be circumvented. Cross-entropy, negative log-likelihood and related next-token objectives cannot alone implement Solomonoff induction: they fit supplied conditionals rather than a program-weighted universal mixture. More computation can improve fit within a fixed objective but cannot change its inductive principle without external hyperparameter or architectural tuning; they alone do not deliver Solomonoff-Levin optimal prediction. For finite learners and observers, theoretical boundaries become less decisive and approaches diverge. Resource-bounded estimators are finite mechanism-search tools whose divergence does not violate algorithmic information conservation. Twenty-six served language-model checkpoints across five pre-training families, 0.8-35 billion parameters and 1.9-8.5 bits per weight, read as next-digit predictors on in-context streams of a declared class of 177,591 programs, put mass on continuations the class never produces, and none stays within the 3.32 bits by which a mixture carrying a Krichevsky-Trofimov code at weight 0.10 can trail that code: at twenty-four symbols the best trails it by 4.5 bits, the median by 36.5 and the worst by 128. Over 256 symbols the three checkpoints nearest that bound settle to 7-13 bits, as a mixture carrying the code at a smaller weight would, while others keep losing ground; and the reading closest to Solomonoff's allocation tracks the mass the models put outside the alphabet, so neither scale nor post-training is shown to move it towards a program-weighted mixture. Frontier developers adopt neurosymbolic approaches, including Fable and Astra, incorporating model synthesis via neurosymbolic computation. They are no longer purely statistical LLMs, making them better, though still limited, candidates for higher forms of induction and model synthesis. Our results contradict recent literature suggesting LLMs are Solomonoff-induction estimators based on arguments related to purely statistical compression algorithms.
\end{abstract}


\section{Introduction}

The modern idea of an intelligence explosion begins with a positive-feedback hypothesis. Good's ``ultraintelligent machine'' was defined by the possibility that a machine capable of improving machine design could initiate a process in which better systems become increasingly effective at producing still better systems~\cite{good1966}. Vinge later placed such a transition at the centre of the technological Singularity, and Kurzweil developed a broader account in which accelerating technological change eventually produces intelligence beyond ordinary human forecasting~\cite{vinge1993,kurzweil2005}. These accounts differ substantially, but they share one structural premise: improvement must feed back into the process that produces further improvement. LeCun's JEPA~\cite{jepa} and Hutter's AIXI~\cite{hutter2005} are other versions of and approaches to AGI frameworks based on planning, abstraction, and Solomonoff induction whose instantiation often involves aspects of Minimum Description Length techniques that fall back into statistical compression and Shannon-entropy estimators with salient limitations.

The recent success of generative models makes one possible feedback mechanism look unusually simple. A capable model can generate text, code, images, evaluations and synthetic training examples at enormous scale. It can assist in constructing its own training corpus and, increasingly, in writing or modifying parts of the software around it. This invites a tempting identification: if generation can be automated and the generated material can train a successor, perhaps recursive self-training is already the beginning of recursive self-improvement.

A parallel theoretical question has become explicit at the intersection of Algorithmic Information Theory (AIT) and machine learning: whether neural sequence predictors, and Transformers or LLMs in particular, should be understood as approximations to Solomonoff induction. Grau-Moya et al. meta-train neural predictors on sequences generated by Universal Turing Machines in an attempt to amortise a universal predictor into neural networks~\cite{graumoya2024}. Young and Witbrock propose modelling Transformers as bounded approximations to Solomonoff induction~\cite{youngwitbrock2024}. Wan and Mei make the stronger claim that standard LLM loss minimisation computationally approximates the Solomonoff prior and that next-token prediction implements an approximate form of Solomonoff induction~\cite{wanmei2025}. These works make the question precise enough to be tested rather than treated as analogy. The present paper asks which properties follow from the standard likelihood objective itself, which require additional inductive structure, and what changes when program-weighted mechanism search is introduced explicitly. Because the question is now testable, we do not leave it at the level of analysis. Section~\ref{sec:servedllm} reports direct measurements on twenty-six served checkpoints, from the companion study~\cite{ozelimzenil2026b}. There the guarantee that makes a mixture worth its cost, bounded regret against a complexity-blind code, is measured at finite horizon: no checkpoint holds the constant of a mixture that carries that code at weight $0.10$, the checkpoints nearest it settle to a constant two to four times larger over the horizon read, and the others keep losing ground.

Wan and Mei's later formulation describes the next-token connection as a computable surrogate and makes explicit the assumption that the constructed surrogate prior is already close to the Solomonoff prior~\cite{wanmei2026}. The apparent tension with the result developed here is resolved by separating two levels of analysis: what a likelihood objective entails by itself, and what becomes possible when a trained likelihood model is embedded in an additional program-based coding construction.

However, generating more samples from a model does not by itself define what would make a successor better. If the successor is trained only to reproduce the predecessor's distribution, exact optimisation reproduces that distribution. If finite sampling, filtering or optimisation moves the model away from its predecessor, the movement has no automatic interpretation as improvement. It may be useful, neutral or destructive. A positive-feedback theory therefore needs more than recursion. It needs a directional criterion.

The model-collapse literature, where model collapse denotes degenerative loss or distortion arising across generations of recursively trained generative models, is important because it makes one failure mode of recursive generation visible. Replacement with synthetic data can lose rare events, amplify distortions and change scaling behaviour, while accumulation, replay, correction and other workflows can behave very differently~\cite{alemohammad2024,shumailov2024,dohmatob2024,fu2024,gerstgrasser2024,gillman2024,kazdan2025,suresh2025,zhu2025}. The lesson is not that synthetic data are intrinsically destructive. It is that the update operator matters. Recursive learning is not one process, and the proportion of synthetic data is not enough to determine its asymptotic behaviour.

The broader issue is rather epistemic. An intelligence explosion is normally imagined as more than a sequence of models that become increasingly self-consistent. The system must somehow improve its representation of the world, its ability to intervene successfully, its algorithms, its scientific theories, its designs or its own architecture. Those improvements require criteria that distinguish a better mechanism from a merely different or more internally coherent one. In science those criteria are supplied by observation, experiment, invariance, prediction and reproducibility. In formal domains they can be supplied by axioms, proof checkers and executable specifications. In engineering they may come from measured performance under independently defined constraints.

This work develops that point while correcting two formulations that are too strong. A per-generation grounding fraction need not remain bounded away from zero, because a vanishing fraction can still exert unlimited cumulative correction. Algorithmic priors, which favour shorter effective descriptions, also do not escape the data-processing inequality, which limits mutual information under post-processing, or Levin's algorithmic-information non-growth results, which bound information growth under effective transformations. Their practical relevance lies elsewhere: under finite resources, different representations and search procedures can make different parts of the same available information operationally accessible. Formal information conservation sets a ceiling; it does not choose an empirical method.

The central thesis is therefore narrower than a universal impossibility theorem, but broader than a model-collapse calculation. Closed-loop distribution matching is not itself an improvement operator. The paper also separates the claim that an LLM may learn algorithmic regularities or even implement bounded program-like procedures from the stronger claim that likelihood training itself implements Solomonoff induction. A credible recursive self-improvement architecture must contain a process that proposes candidate changes and evaluates them against criteria not reducible to reproducing the current output distribution. This paper calls that functional requirement \emph{symbolic model synthesis}: the construction, manipulation and testing of explicit generative structures such as programs, equations, causal models, formal rules or executable mechanisms. The term is functional rather than historical. A neural system that genuinely performs these operations counts as a model-synthesis system in the sense intended here.

The distinction is easiest to see by returning to the classical intelligence-explosion argument. Good did not merely imagine a machine producing more artefacts of the same kind. His argument concerned a machine able to improve the design of intelligent machines, including eventually itself. The feedback is therefore second-order: improved capability must improve the process by which further capability is produced. Vinge's Singularity inherits this recursive structure, while Kurzweil embeds it in a wider argument about accelerating technological change. None of these accounts is reducible to the statement that a sufficiently capable model can manufacture an arbitrarily large training set. The relevant question is whether the system has a reliable operator that maps its present state into a demonstrably better successor and whether improvement of the system also improves that operator.

This distinction matters because several quite different activities are now described informally as ``self-improvement''. A language model can generate synthetic examples, critique its own answers, write code, propose changes to prompts, search over tool configurations, or help train a later model. Each may be useful, but none is by definition an improvement step. A generated modification becomes an improvement only relative to some criterion, and the criterion itself must remain meaningful as the system changes. If the same model family supplies the proposal, the target and the judgement, a positive score can reflect increasing internal agreement rather than increasing competence with respect to the world or to an independently specified task.

There is therefore an important asymmetry between candidate production and candidate validation. Modern generative models have made candidate production extraordinarily cheap. They can propose code, proofs, explanations, molecular structures, experimental protocols and model architectures at a scale that was previously impossible. This changes the economics of search, but it makes validation more, not less, important. Once proposals become abundant, the bottleneck moves towards deciding which proposals survive contact with constraints that are not simply restatements of the generator's current preferences. A theory of recursive self-improvement has to account for this second half of the loop.

The point can also be expressed in terms of error correction. Human scientific and engineering institutions are not reliable because every individual proposal is well grounded. They are reliable, when they are reliable, because conjectures can be rejected by measurements, failed replications, incompatible observations, formal contradictions or engineering tests. An autonomous machine need not reproduce those institutions, but it needs some functional analogue if it is to improve open-endedly rather than merely become more self-consistent. The central issue is thus not whether humans remain present, but whether the system retains channels through which its own current model can be proved wrong.

\section{Preliminaries and notation}

\emph{Model collapse} refers here to degenerative loss or distortion that can arise in recursively trained generative models, including loss of support or diversity under replacement-style workflows; the precise behaviour depends on the update rule and data-retention scheme~\cite{alemohammad2024,shumailov2024,dohmatob2024,gerstgrasser2024}. \emph{Grounding} denotes corrective influence supplied by information or constraints not reducible to neutral reproduction of the current model distribution. In the simple mixture model below it is represented by an external data weight, but elsewhere it can also arise from retained evidence, environmental interaction, experiments, formal verification or invariant task criteria. We use the term \emph{epistemic openness} for the broader architectural property of remaining answerable to such non-circular correction.

The statistical notation is standard~\cite{coverthomas2006}. Upper-case letters such as $P$, $Q_t$, and $R$ denote probability distributions unless otherwise stated. $P$ will usually denote a fixed external or reference distribution, $Q_t$ the model distribution at recursive step $t$, and $R$ a candidate successor distribution. Lower-case $q_t$ denotes a probability vector on a finite alphabet. The quantity $N_t$ denotes the sample size used at step $t$. The parameter $\alpha_t\in[0,1]$ denotes the weight assigned to the external reference distribution in the simple grounded mixture model, and $\beta_t=\prod_{s<t}(1-\alpha_s)$ denotes the residual multiplicative influence of the initial discrepancy after $t$ updates. For a discrete distribution $q$, Shannon entropy is $H(q)=-\sum_i q_i\log q_i$, while $D_{\mathrm{KL}}(p\Vert q)$ denotes Kullback--Leibler divergence. Mutual information between random variables $X$ and $Y$ is denoted by $I(X;Y)$. The base of the logarithm is immaterial to the qualitative arguments; base $2$ is used when information is interpreted in bits.

An autoregressive LLM represents a token sequence $x_{1:n}$ by the conditional factorisation
\begin{equation}
p_\theta(x_{1:n})=\prod_{t=1}^{n}p_\theta(x_t\mid x_{<t}),
\label{eq:autoregressive}
\end{equation}
where $x_{<t}$ is the preceding context and $\theta$ denotes the model parameters~\cite{vaswani2017,brown2020}. Thus the probability assigned to a complete sequence is the product of the probabilities assigned to its successive tokens conditional on the preceding context, which is the standard autoregressive next-token factorisation. For a finite vocabulary $\mathcal V$, a decoder produces a real-valued logit $z_\theta(v\mid x_{<t})$ for each token $v\in\mathcal V$, and a softmax converts these logits into a conditional next-token distribution,
\begin{equation}
p_\theta(v\mid x_{<t})=
\frac{\exp z_\theta(v\mid x_{<t})}
{\sum_{u\in\mathcal V}\exp z_\theta(u\mid x_{<t})}.
\label{eq:softmaxnext}
\end{equation}
The normalisation maps the unconstrained logits to non-negative probabilities that sum to one across the vocabulary, with larger logits receiving a larger share of the next-token mass. Generation then selects or samples the next token from this conditional distribution. Standard autoregressive pre-training commonly minimises token-level negative log-likelihood, equivalently cross-entropy or maximum-likelihood loss,
\begin{equation}
\mathcal L_{\mathrm{NLL}}(\theta)
=-\mathbb E_{X\sim P}\left[\sum_t\log p_\theta(X_t\mid X_{<t})\right].
\label{eq:tokenloss}
\end{equation}
The loss is large when the model assigns low probability to the token that actually occurs, so minimisation rewards progressively better agreement with the conditional token distribution represented in the training data. Perplexity is an exponential transformation of average negative log-likelihood and therefore preserves the same model ordering for a fixed evaluation set. We refer below to cross-entropy, negative log-likelihood, maximum likelihood and their direct equivalents as \emph{likelihood-based next-token objectives}. These definitions concern the predictive interface and optimisation criterion; they do not assume that a neural network is incapable of learning an internal algorithm or structured representation.

The \emph{data-processing inequality} (DPI) states that if random variables form a Markov chain $X\to Y\to Z$, then $I(X;Z)\leq I(X;Y)$~\cite{coverthomas2006}. In this paper DPI is used only as an information-conservation constraint: post-processing cannot create additional mutual information about an external variable beyond that contained in the inputs to the processing chain. Later, $M$ denotes an unknown mechanism, $D$ the observations available to a learner, $B$ fixed background information and $Y$ the learner's output. The paper also introduces $T_b$ for the transcript obtainable under a finite inspection budget $b$, $\mathcal A_b$ for a class of procedures available under computational budget $b$, and $I_b^{\mathrm{acc}}$ for the corresponding task-relevant accessible information. The last quantity is an operational construction introduced here, not a replacement for standard mutual information.

The algorithmic-information terminology follows the standard framework of Solomonoff induction, Kolmogorov complexity and Levin's algorithmic probability~\cite{solomonoff1964,levin1974,li2019,hutter2005}. Let $U$ be a fixed universal prefix machine and $p$ a self-delimiting program . The prefix Kolmogorov complexity of a finite object $x$ is $K_U(x)=\min\{|p|:U(p)=x\}$, defined up to the usual additive constant associated with the choice of universal machine. Algorithmic probability assigns greater weight to outputs generated by shorter program s; schematically, $m_U(x)=\sum_{p:U(p)=x}2^{-|p|}$. The coding theorem relates these quantities by $K_U(x)=-\log_2 m_U(x)+O(1)$. Exact Kolmogorov complexity and the universal semimeasure are uncomputable. Levin's information non-growth results are algorithmic counterparts of the same conservation intuition: under the conditions of the theorem, effective processing cannot systematically manufacture unbounded algorithmic information that was absent from the relevant inputs~\cite{levin1974,li2019}. Nothing in this paper treats these results as false or violated; the issue examined is what finite procedures can make operationally accessible under bounded resources.

The \emph{Coding Theorem Method} (CTM) is a finite computational method that estimates algorithmic probability and hence algorithmic complexity by enumerating or systematically running classes of small programs or machines and measuring their output frequencies~\cite{delahayezenil2012}. CTM is therefore an approximation within a specified finite machine and runtime regime based on or inspired by the universal distribution.

The \emph{Block Decomposition Method} (BDM) is a greedy algorithm that extends CTM-style estimates to larger objects by decomposing them into smaller components, reusing local algorithmic-complexity estimates and accounting for repeated components~\cite{zenil2018} penalising by repetitions according to classical information theory hence combining the best of both classical and algorithmic worlds. Its estimates consequently depend to some controlled degree on the chosen decomposition~\cite{zenil2018} as well as on the underlying CTM values for a reference machine, shown to in general produce convergent values (hence possibly not much dependant on reference machine as long as it is `natural')~\cite{delahayezenil2012} and, in the worse case, converging to Shannon entropy rate in the limit~\cite{zenil2018}.

\emph{Algorithmic Information Dynamics} (AID) studies changes in estimated algorithmic information under controlled perturbations in order to analyse structural contribution, dynamical organisation and reprogram mability~\cite{zenil2019,zenil2023}. AID is used here as a perturbational and mechanism-oriented framework, not as a claim that algorithmic complexity alone identifies causal truth.

A \emph{generative mechanism} means a program, equation, causal model, simulator or other rule capable of producing or explaining the observations and of entailing consequences beyond simple replay of the observed sample. Different mechanisms may be observationally compatible, so observational fit alone need not identify a unique causal structure; causal identification requires additional assumptions, interventions or other identifying information depending on the problem~\cite{pearl2009}. The term \emph{symbolic model synthesis} is used in a functional sense for constructing, manipulating and testing explicit generative structures whose consequences can be checked against evidence or formal constraints. It does not require classical symbolic AI or a separate symbolic module. A neural or hybrid system that performs these functions counts as a model-synthesis system in the sense used here. Large language model (LLM), artificial general intelligence (AGI) and artificial superintelligence (ASI) are used in their conventional broad senses; no particular operational definition of AGI or ASI is required by the mathematical results below.

\section{Recursive self-improvement and the autonomy premise}

The Singularity argument is often described in terms of autonomy, but autonomy has several meanings that should not be conflated. A system may become operationally autonomous because it needs less direct human labour. It may become computationally autonomous because it writes code, allocates resources or schedules its own searches. Neither form of autonomy requires epistemic isolation. An autonomous scientific agent could continuously read instruments, query databases, run experiments, interact with an environment and submit conjectures to independent verifiers. Such a system may require little human intervention while remaining strongly grounded in information that does not originate from its own current model distribution.

The relevant distinction is therefore between \emph{autonomy} and \emph{endogenisation}. A recursive loop becomes epistemically closed to the extent that its training targets, evaluators and evidence are generated from the same endogenous model family without an independent correctness signal. This is the regime in which self-consistency can increasingly substitute for external correction. It is also the regime that makes model collapse relevant to arguments about recursive self-improvement. The problem is not that humans have left the loop. The problem is that the world, an invariant evaluator or another non-circular source of constraint has effectively left the loop.

This distinction changes how the original grounding parameter should be interpreted. If $\alpha_t$ measures the proportion of samples drawn directly from an external distribution, then a vanishing $\alpha_t$ does not by itself imply loss of grounding. The absolute amount of new evidence may still grow, old evidence may be retained, or a vanishing proportion may have sufficient cumulative influence. Moreover, some forms of grounding cannot naturally be represented as a sample fraction at all. A proof checker, a conservation law, an executable specification or an experimental outcome can constrain an update even when almost all training tokens are synthetic.

It is useful to distinguish three generic loops. In a closed distributional loop, the current model generates the examples and the successor is trained primarily to match the resulting distribution. In an externally evaluated loop, the model may still generate most candidate data or actions, but a rule set, environment, benchmark, verifier or measurement supplies an independent selection criterion. In a mechanism-seeking loop, the system constructs an explicit generative hypothesis, derives consequences that were not simply replayed from the training sample, and uses observation or formal checking to discriminate among hypotheses. These loops can share the same neural components while having very different epistemic structures.

Self-play illustrates why the distinction matters. A chess or Go system can generate essentially all of its own game trajectories, yet the rules and the win condition remain fixed outside the current policy distribution. Recursive generation is therefore not closed in the relevant sense. Likewise, code-generation systems can improve measured runtime if execution on a trusted benchmark supplies the score, and theorem-proving systems can generate their own conjectures and candidate proofs if a sound checker supplies correctness. These examples show that synthetic generation is compatible with improvement when something independent decides what counts as success.

The difficult case for broad recursive self-improvement is that the evaluator itself may need to improve. A fixed chess score defines a bounded task. An intelligence explosion, by contrast, is usually taken to involve expanding competence across domains, improving representations, discovering new objectives or subgoals, and modifying the machinery that performs evaluation. The more of this stack becomes endogenous, the more important it becomes to ask what prevents the system from optimising its own proxy, narrowing its hypothesis space or mistaking internal agreement for external correctness.

This is the reason the paper retains the language of the Singularity. The issue is not whether a future system can be autonomous. It is whether the positive-feedback loop assumed by Good, Vinge and Kurzweil has been specified at the epistemic level. Recursive generation provides an abundant source of candidate material. It does not, on its own, supply the criterion that turns recursive change into recursive improvement.

This way of framing autonomy also clarifies why self-play and formal reasoning are not counterexamples to the argument. In self-play, most of the trajectories may be generated internally, but the game rules and terminal conditions provide a stable external structure. In formal mathematics, a model may generate both conjectures and candidate proofs, but a proof checker can reject an invalid derivation without consulting the model's confidence. In program  optimisation, the model can propose its own rewrites while execution time, memory use or test-suite correctness provide measurements that are not determined by the proposal distribution. The proportion of human-authored material can approach zero in all three cases without producing epistemic closure.

Open-ended scientific discovery is harder because the relevant criteria are not always fixed in advance. A system investigating an unfamiliar physical process may need to decide which variables to measure, which instruments to build, which interventions distinguish competing mechanisms and which apparent regularities are artefacts of measurement. Here improvement cannot be reduced to maximising a static benchmark. The learner must participate in constructing the very tests that make its hypotheses answerable to the world. This is one reason the paper treats model synthesis and independent evaluation as coupled operations rather than as two permanently separate modules.

The same distinction prevents an overly simple reading of ``external grounding''. External information is not synonymous with human text. A sensor reading, a failed experiment, a theorem checker, an executable simulator constrained by known physics, a database measurement or an interaction with another agent can all alter the evidential situation. Conversely, a large corpus of nominally real data can provide little corrective value if it is redundant with what the model already represents. The mathematically convenient parameter $\alpha_t$ captures one kind of mixture process, but epistemic openness is a broader architectural property.

\section{Recursive self-training: identity, grounding and collapse}
\label{sec:recursive}

Generative models are naturally described by probability distributions, whether explicitly or implicitly. An autoregressive language model assigns conditional probabilities to token sequences; a diffusion model defines a generative process through iterative denoising; a GAN defines a distribution through the output of its generator. These architectures differ, but the recursive-training question can be isolated at a higher level: what happens when the distribution produced by one generation increasingly determines the training target of the next?

Let $Q$ be a model distribution and suppose a successor distribution $R$ is trained, in the population limit, by minimising log-loss on samples from $Q$. The expected loss is

\begin{equation}
\mathbb{E}_{X\sim Q}[-\log R(X)]
=
H(Q)+D_{\mathrm{KL}}(Q\Vert R).
\end{equation}
Here $H(Q)$ is fixed by the target distribution, whereas the KL-divergence term measures the discrepancy between the successor $R$ and $Q$ and is the only part that can be reduced by changing $R$. Since this term is non-negative and vanishes when $R=Q$, exact self-training has the optimum

\begin{equation}
R^{*}=Q.
\end{equation}
At the population level, the best possible successor under this objective is therefore the distribution it is trained to imitate: pure self-imitation is an identity operation expressing fidelity rather than an intrinsic direction of improvement. This observation is more important for recursive self-improvement than a universal collapse theorem would be. If the objective is only to reproduce $Q_t$, the exact optimum is $Q_t$. If another objective moves a successor in a useful direction, that objective has entered the effective learning system and should be analysed as part of the loop.

Grounding can be represented in a simple population model by mixing a fixed reference distribution $P$ with the current model distribution $Q_t$:

\begin{equation}
Q_{t+1}
=
\alpha_t P+(1-\alpha_t)Q_t,
\qquad
0\leq\alpha_t\leq1.
\label{eq:mix}
\end{equation}
Here $\alpha_t$ is the fraction of the update attributable to the external reference distribution at generation $t$, while $1-\alpha_t$ is the retained contribution from the current model. The update is therefore a weighted average of external correction and self-retention. Subtracting $P$ and iterating gives

\begin{equation}
Q_t-P
=
\beta_t(Q_0-P),
\qquad
\beta_t
=
\prod_{s=0}^{t-1}(1-\alpha_s).
\label{eq:beta}
\end{equation}
Thus the discrepancy from $P$ retains its direction but is scaled by $\beta_t$, which accumulates every preceding retention factor. The asymptotic behaviour is therefore governed by the whole grounding schedule through the cumulative product $\beta_t$, not by the pointwise limit of $\alpha_t$. If no $\alpha_t$ is exactly one, $\beta_t$ tends to zero when the cumulative grounding is sufficient. Under the usual infinite-product criterion, this occurs when

\begin{equation}
\sum_{t=0}^{\infty}\alpha_t=\infty.
\label{eq:alphasum}
\end{equation}
The condition allows individual corrections to become arbitrarily small while requiring their cumulative influence not to remain finite; under the stated infinite-product conditions this is sufficient for the residual factor $\beta_t$ to vanish. A grounding fraction can consequently vanish while continuing to remove the initial discrepancy. For example, $\alpha_t=1/(t+2)$ tends to zero but gives $\beta_t=1/(t+1)$, so $Q_t$ converges to $P$. By contrast, $\alpha_t=1/(t+2)^2$ also tends to zero, but the cumulative product remains positive and a residual contribution from $Q_0$ survives. A condition such as $\inf_t\alpha_t>0$ is therefore sufficient in this model but not necessary.

This calculation also clarifies what the grounding parameter represents and, equally importantly, what it does not. The instantaneous value of $\alpha_t$ measures the contribution of the external distribution at a particular stage, but it does not by itself determine whether the process remains grounded. The relevant quantity in this model is the cumulative influence of the sequence $\{\alpha_t\}$, as captured by $\beta_t$. More generally, even this cumulative mixture is only one mathematical representation of grounding. Experiments, environmental interaction, formal verification, invariant constraints and independently defined performance criteria can all provide corrective information without being naturally expressible as a fraction of training samples. The mixture model therefore illustrates a broader progression from instantaneous grounding, to cumulative grounding, and ultimately to epistemic openness as an architectural property. For recursive self-improvement, the important question is not whether $\alpha_t$ remains bounded away from zero, but whether the system preserves channels through which its hypotheses, modifications and evaluators remain answerable to constraints that are not simply regenerated from its own current distribution.

\begin{figure}[ht]
\centering
\includegraphics[width=0.75\textwidth]{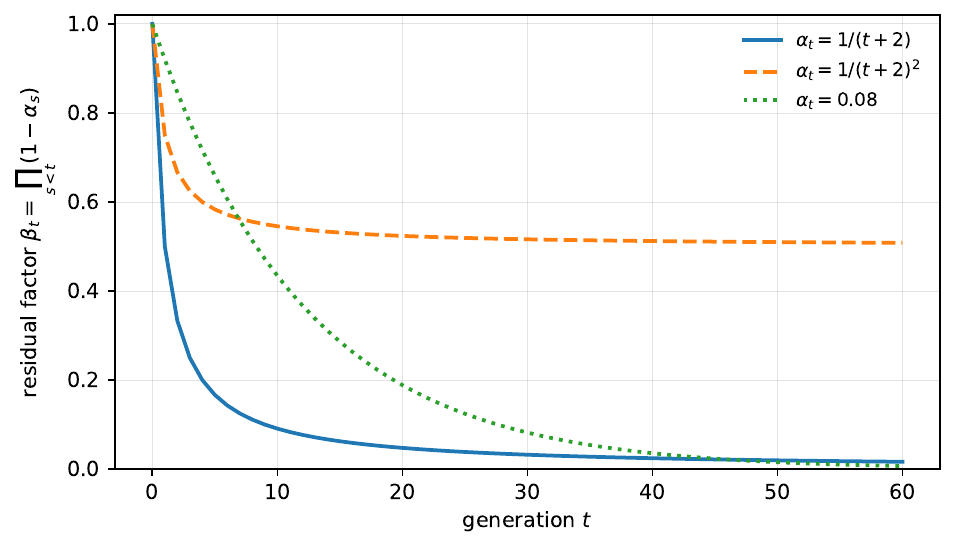}
\caption{Three grounding schedules/regimes and the corresponding residual factor $\beta_t$. A vanishing per-generation fraction does not determine the asymptotic result. The cumulative product does.}
\label{fig:grounding}
\end{figure}

Finite sampling introduces a different effect. On a finite alphabet, let $q_t$ be the current probability vector, draw $N_t$ samples from it, and let the empirical frequencies become $q_{t+1}$. Define the Gini--Simpson diversity by

\begin{equation}
V(q)=1-\sum_i q_i^2.
\end{equation}
This quantity vanishes when all probability mass is concentrated on one outcome and grows as the mass is spread across several outcomes, making it a simple diversity measure for the finite-alphabet model. For multinomial resampling,

\begin{equation}
\mathbb{E}\left[V(q_{t+1})\mid q_t\right]
=
\left(1-\frac{1}{N_t}\right)V(q_t).
\label{eq:diversity}
\end{equation}
Because $1-1/N_t<1$ for every finite sample size, each replacement step removes on average a fraction $1/N_t$ of the current diversity, with the effect diminishing as $N_t$ grows. Repeated finite replacement therefore reduces expected diversity whenever $N_t$ is finite. With a fixed finite sample size, the process eventually concentrates on a random surviving state. The mechanism is straightforward: finite resampling introduces sampling fluctuations, and making the empirical sample the complete target of the next generation repeatedly converts those fluctuations into the new population.

This result is intentionally narrow. It describes neutral multinomial replacement, not neural optimisation in general. Its value is that it isolates an exact source of degeneracy: repeated finite resampling without a restoring target. It also shows why a universal state-independent entropy decrement is too strong. Near a point mass the expected change can be arbitrarily small, and at a point mass it is zero.

Grounding and sampling must therefore be analysed together. In the same finite-alphabet model, a useful sufficient condition for mean-square consistency is

\begin{equation}
\sum_{t=0}^{\infty}\alpha_t=\infty,
\qquad
N_t\alpha_t\longrightarrow\infty.
\label{eq:groundedsufficient}
\end{equation}
The two requirements play distinct roles: the first supplies unbounded cumulative correction towards the reference distribution, while the second makes finite-sampling noise negligible relative to that correction. Together they are sufficient for the mean-square error to vanish in this model, yet neither requires the grounding fraction itself to remain bounded away from zero. For example, the schedule $\alpha_t=1/(t+2)$ and $N_t=(t+2)^2$ satisfies both conditions even though $\alpha_t$ vanishes.

These simple results accord with the broader model-collapse literature once different workflows are kept separate. Recursive replacement can produce collapse or tail loss~\cite{alemohammad2024,shumailov2024,dohmatob2024,suresh2025}; mixed-data analyses can control error under explicit conditions~\cite{fu2024}; correction functions can stabilise self-consuming loops~\cite{gillman2024}; and accumulation or retention can prevent the degradation seen under replacement~\cite{gerstgrasser2024,kazdan2025}. Token-level editing is another example in which the data-generation rule is altered rather than left as pure self-replacement~\cite{zhu2025}. These results are not contradictory. They concern different update operators.

The conclusion required here is consequently more general than any statement about the limiting value of $\alpha_t$. Exact self-imitation has no improvement direction. Finite recursive replacement can lose diversity. Retained evidence, correction, curation and external objectives can change the dynamics. A model that perfectly reproduces its predecessor has achieved fidelity, not improvement; a model that moves away from its predecessor because of sampling noise has achieved change, not necessarily useful novelty; and a model whose successors are selected by evidence or an independent criterion is participating in a different process altogether. For recursive self-improvement, the decisive question is therefore what corrective information and evaluation remain available to the loop, rather than the proportion of synthetic data considered in isolation.

\section{Information conservation and empirical openness}
\label{sec:conservation}

The data-processing inequality is a theorem about random variables under a specified joint distribution. Let $M$ denote an unknown mechanism, $D$ the observations available to a learner, $B$ fixed background information, and $Y$ the output of a procedure using only $(D,B)$ and internal randomness independent of $M$ once $(D,B)$ are given. Then
\begin{equation}
I(M;Y)\leq I(M;D,B).
\label{eq:dpi}
\end{equation}
The bound states that processing the observations and background information cannot make the output $Y$ contain more mutual information about the unknown mechanism $M$ than was present in those inputs. It constrains information flow without prescribing an efficient inference procedure. Nothing in this work contradicts that inequality~\cite{coverthomas2006}. Levin's non-growth results make an analogous point for algorithmic information under computable and certain random transformations~\cite{levin1974,li2019}. A fixed computation cannot manufacture information about an external mechanism from nothing.

The methodological question is different. Real inference is conducted with local and partial access to data, finite computation, finite memory, restricted representations and bounded search. The fact that all procedures lie below the same information ceiling does not tell us which procedure will actually discover a useful regularity or mechanism. Data processing is therefore a foundational starting point, not a method-selection theorem.

A simple example makes the distinction explicit. Let $M$ be uniformly distributed over $n$ positions and let the fixed data object $D$ be a one-hot vector whose single $1$ occurs at position $M$. Globally, $D$ identifies $M$ completely. A learner that can inspect only the first $b$ coordinates has less operational access. If the $1$ is found, the index is known; if it is not found, $M$ remains uniformly distributed over the uninspected positions. The mutual information in the finite query transcript is
\begin{equation}
I(M;T_b)=\log_2 n-\frac{n-b}{n}\log_2(n-b).
\label{eq:localaccess}
\end{equation}
For this example the expression quantifies exactly how much the learner has discovered after inspecting $b$ coordinates: it is zero at $b=0$, increases with the inspection budget, and reaches the full $\log_2 n$ bits when all positions are accessible, even though the underlying data object and the global information ceiling never change. No information law is violated. More of the fixed information has become operationally accessible to the bounded procedure.

The same idea can be stated more generally. Let $\mathcal A_b$ be a nested class of task-relevant procedures available under computational budget $b$, with outputs restricted to the kind of prediction, mechanism description or decision under study. Define
\begin{equation}
I_b^{\mathrm{acc}}(M;D,B)
=\sup_{A\in\mathcal A_b} I\bigl(M;A(D,B)\bigr).
\label{eq:accessible}
\end{equation}
The supremum represents the best task-relevant information about $M$ that any procedure available within budget $b$ can extract from the fixed evidence $(D,B)$. If $\mathcal A_b\subseteq\mathcal A_{b+1}$, then
\begin{equation}
I_b^{\mathrm{acc}}(M;D,B)
\leq I_{b+1}^{\mathrm{acc}}(M;D,B)
\leq I(M;D,B).
\label{eq:accessiblebound}
\end{equation}
The inequalities make the resource distinction explicit. Enlarging the available procedure class cannot reduce the best accessible information, but no such procedure can exceed the information contained in the complete inputs; additional computation can therefore improve access without violating the global ceiling. This is not proposed as a new universal information measure. It is an operational device for separating a global information ceiling from what a specified class of finite procedures can actually make usable for a task.

This distinction is especially important when applying Levin's theorem to open-ended algorithmic search. An entire schedule of increasingly large searches can indeed be described by one meta-algorithm, and information non-growth applies to the whole computation. But describing an unlimited search does not make the result of an unperformed search available at an earlier finite stage. Computability in principle and accessibility under a finite resource budget are different methodological facts. The theory bounds information; it does not erase computational depth.

An empirical system can therefore remain open-ended in two different senses. Changing representation, search depth or hypothesis class can increase computational access to consequences already implicit in fixed data. Observation, intervention or independent verification can instead supply genuinely new empirical input. The former changes accessibility without changing the global information in the original observations; the latter changes the information set itself. Scientific inference uses both.

This is the intended position of CTM relative to DPI and Levin non-growth. CTM does not escape either theorem, and a universal prior is not a permanent stream of empirical grounding about the particular mechanism under study. The claim is methodological: two learners may satisfy the same information-conservation bound yet differ substantially in what structure they can expose, manipulate and test under finite resources. Information conservation remains true while being largely non-discriminating between empirical methods operating below the same ceiling.

The information-theoretic issue is sometimes made to carry more methodological weight than it can support. The data-processing inequality answers a precise question about information flow under a specified probabilistic dependence. Levin-style non-growth results answer a related algorithmic question about the inability of effective processing to manufacture unlimited algorithmic information from nothing. Both are indispensable boundary conditions. Neither says that all computable procedures presented with the same finite observations are equally useful, equally efficient or equally capable of finding a compact explanation within a finite time.

This distinction is familiar in ordinary computation. The fact that the output of a program  is determined by its input does not imply that the output is already available to a bounded observer in any operational sense. Factoring a large integer, solving a difficult search problem or running a long simulation may reveal consequences of a fixed input that were always implicit in it, yet obtaining those consequences can require substantial computation. In empirical inference the situation is richer because the choice of representation also determines which regularities are cheap to express and search. A bound on total information does not erase these differences in computational accessibility.

For this reason, ``open-ended'' has two meanings in the present argument. A learner can be computationally open-ended by enlarging the class of hypotheses it can search, the depth of programs it can execute, or the representations it can construct. This can expose consequences already implicit in its existing evidence. A learner can also be empirically open by acquiring new measurements, interventions or independently verified outcomes. The first changes access to information without violating conservation; the second changes the evidential input itself. Scientific practice continually alternates between the two. A theory suggests a calculation or experiment, the result changes the evidence, and the revised evidence motivates another theory.

This is also why a universal prior should not be described as though it were a permanent stream of observations about the particular world being studied. Its role is structural. It changes the ordering of candidate explanations by favouring shorter generative descriptions. Whether that bias is useful depends on the relation between the chosen description language, the available computational budget and the actual mechanism. The empirical advantage, when there is one, lies in reaching a useful hypothesis with finite resources, not in violating a global information bound.

The distinction has a direct methodological consequence. An objection based only on DPI or non-growth cannot decide between a distributional estimator, a causal-search procedure, a theorem prover and a program -synthesis system if all operate on the same evidence. They all respect the relevant conservation law. What matters empirically is which consequences they can derive, which hypotheses they can represent, how much computation they require and whether their outputs survive independent tests. The information bound is the floor of the discussion, not its conclusion.

\section{Mechanism discovery, causality and algorithmic model synthesis}

The original motivation for introducing symbolic and algorithmic methods was to distinguish distribution fitting from mechanism discovery. That distinction should not be stated as a claim that neural networks are incapable of representing mechanisms, nor as a claim that a particular loss function is intrinsically incapable of causal reasoning. Neural systems can implement algorithms, learn structured latent variables, call tools and participate in experimental design. The narrower point is that fitting an observational distribution does not, by itself, identify which causal or generative mechanism produced it. Different mechanisms can induce the same observational distribution, and causal conclusions require additional assumptions, interventions or other identifying structure~\cite{pearl2009}.

This matters for recursive self-improvement because prediction and explanation can come apart. A model can reproduce the statistical regularities of its training environment extremely well while remaining unable to distinguish among mechanisms that agree on observed data but diverge under intervention, distribution shift or extrapolation. Conversely, a compact mechanism can make predictions outside the observed support, but only if the mechanism has been identified well enough to justify those predictions. The issue is not correlation versus causation as a slogan. It is the logical gap between matching observations and selecting among observationally compatible generators.

Kant's distinction between analytic and synthetic judgements provides a useful historical analogy, but only an analogy~\cite{kant1998}. The paper does not claim that LLMs are literally confined to Kantian analytic judgements, nor that a neural network cannot generate a novel proposition. The relevant contrast is between rearranging or extending consequences within an accepted representational frame and acquiring a warranted new model of how the world works. For the purposes of recursive self-improvement, what matters is not novelty of wording but whether a proposed law, mechanism or design survives a test whose outcome is not fixed by the model's own current beliefs.

Algorithmic probability offers one principled way to organise a search over generative explanations. For a universal prefix machine $U$, the Solomonoff--Levin semimeasure can be written schematically as
\begin{equation}
m(x)=\sum_{p:U(p)=x}2^{-|p|},
\end{equation}
where each program $p$ that generates $x$ contributes weight $2^{-|p|}$, so shorter generating programs contribute exponentially more mass. Solomonoff--Levin induction thereby organises prediction around a mixture of possible generative mechanisms rather than observed frequency alone, with the coding theorem linking high algorithmic probability to low Kolmogorov complexity up to the usual machine-dependent additive constant~\cite{solomonoff1964,levin1974,li2019,hutter2005}. Exact Kolmogorov complexity and the full universal distribution are uncomputable, so every practical implementation is necessarily finite and approximate.

The motivation for introducing algorithmic probability is not that statistical regularity is unimportant. It is that statistical similarity and generative simplicity answer different questions. Two observations can be far apart under a surface metric and nevertheless be generated by the same short program ; conversely, a distribution can be fitted accurately by many mechanisms that disagree sharply outside the observed support. A generative description is valuable when it compresses the observations by specifying a process that can be run, modified and interrogated.

Solomonoff's construction provides an idealised formal version of this preference. programs that generate an observation contribute probability mass weighted by their description length. The resulting universal distribution is not computable, and this limitation is fundamental rather than an implementation inconvenience. Its relevance here is therefore methodological: it defines a principled direction for searching over generative explanations, while practical methods must approximate that direction within finite program  classes and runtime budgets.

The distinction becomes especially concrete at the next-token interface. For a fixed data-generating distribution, expected token log-loss is minimised by the corresponding conditional data distribution whenever that distribution is representable by the model. This is the conditional analogue of the identity calculation in the previous section: cross-entropy can be written as entropy plus a KL divergence, and its optimum is obtained by matching the target conditional distribution. Negative log-likelihood and maximum likelihood express the same optimisation, while perplexity is a monotone transformation of the resulting average log-loss. These objectives can produce extremely capable predictors, but the objective itself contains no term that enumerates candidate generating programs or weights them by $2^{-|p|}$.

This yields a limited but sharp conclusion. A standard autoregressive LLM trained by cross-entropy or a cognate likelihood-based next-token objective is not, by virtue of that objective alone, a Solomonoff predictor. The softmax in \eqref{eq:softmaxnext} normalises scores over vocabulary tokens, whereas Solomonoff induction assigns predictive weight through a mixture over programs that can generate compatible continuations. Low next-token log-loss therefore does not imply that the predictor is estimating the universal algorithmic semimeasure. This is not a representational impossibility theorem for transformers: a neural network could in principle learn or implement a Solomonoff-like bounded procedure. The point is that such an algorithmic prior or program -search operation is additional structure, not something entailed or identified by the likelihood objective itself.

\subsection{Solomonoff induction and the location of the inductive burden}
\label{sec:solomonoff}

The distinction above bears directly on recent attempts to interpret neural predictors as approximations to Solomonoff induction~\cite{graumoya2024,youngwitbrock2024,wanmei2025}. The relevant issue is not whether a neural architecture can represent algorithmic procedures. It is where the structural assumptions that determine successful induction enter the effective learning system. Cross-entropy is a scoring rule rather than an architecture, but the estimator obtained by minimising it depends on the hypothesis class, representation and effective probability mass presented during training. A useful way to expose this dependence is to formalise curation as a transformation of the training distribution.

For simplicity, let $P$ be a discrete observational distribution and let $w(x)\geq0$ represent selection, filtering, oversampling, curriculum weighting or another transformation of the probability mass supplied to the learner. Provided $0<Z_w<\infty$, define
\begin{equation}
P_w(x)=\frac{w(x)P(x)}{Z_w},
\qquad
Z_w=\mathbb E_{X\sim P}[w(X)].
\label{eq:curateddistribution}
\end{equation}
Here $w(x)$ changes the relative mass assigned to examples and $Z_w$ renormalises those weights, so $P_w$ remains a probability distribution. A likelihood-based estimator trained on this effective distribution minimises
\begin{equation}
\mathcal L_w(Q)
=
\mathbb E_{X\sim P_w}[-\log Q(X)]
=
H(P_w)+D_{\mathrm{KL}}(P_w\Vert Q).
\label{eq:curatedloss}
\end{equation}
The decomposition is the usual cross-entropy identity applied to the curated distribution $P_w$: once the effective training mass has been reshaped, the likelihood objective rewards agreement with that reshaped distribution. In an unrestricted population model the optimum is therefore
\begin{equation}
Q_w^{*}=P_w.
\label{eq:curatedoptimum}
\end{equation}
With no restriction on the model family, the minimum is reached by reproducing $P_w$ exactly. The likelihood objective therefore inherits the curation embodied in $w$ rather than determining for itself how the original distribution should have been reweighted. In particular, the loss does not determine the weighting function $w$. If an external process changes which observations are retained, how frequently they are sampled, or which synthetic examples are admitted, it has changed the target distribution whose probability mass the estimator is asked to reproduce. This may be entirely legitimate, but the structural work performed by that process should be counted as part of the effective learning architecture rather than attributed to cross-entropy itself.

Architecture and representation can be made explicit by letting $\mathcal Q_A$ denote the family of distributions representable under an architecture or hypothesis class $A$. When the minimum is attained, a population solution is a KL projection
\begin{equation}
Q^{*}_{A,w}
\in
\arg\min_{Q\in\mathcal Q_A}
D_{\mathrm{KL}}(P_w\Vert Q).
\label{eq:architectureprojection}
\end{equation}
When the architecture can represent only distributions in $\mathcal Q_A$, optimisation selects a member of that class minimising the KL discrepancy from $P_w$. The realised estimator is therefore jointly determined by the observational distribution $P$, its external transformation $w$, and the representational class $\mathcal Q_A$. A sufficiently expressive likelihood-trained system can exhibit sophisticated inductive behaviour, but the objective alone still specifies neither the architecture nor how the training mass is constructed. The Universal-Turing-Machine data used by Grau-Moya et al. to meta-learn universal prediction strategies is a particularly direct illustration: the resulting neural learner is trained by ordinary optimisation, while a program-generating process supplies the algorithmically structured meta-training distribution~\cite{graumoya2024}.

One can quantify the magnitude of a distributional intervention, without pretending to measure human effort or semantic quality, by
\begin{equation}
C_{\mathrm{mass}}(w)
=
D_{\mathrm{KL}}(P_w\Vert P).
\label{eq:massintervention}
\end{equation}
Here the KL divergence records only how far the effective training distribution has been displaced from the observational distribution. It is deliberately not a universal measure of curation cost, human labour, intelligence or semantic quality. Its purpose is to separate improvement caused by changing the estimator from improvement caused by changing the probability mass that the estimator is asked to fit.

The role of computational power can be separated in the same way. Suppose additional computation is used only to solve a fixed likelihood-learning problem more accurately. If $\widehat Q_{A,w,c}$ denotes the result obtained under computational budget $c$, its optimisation error can be represented schematically by
\begin{equation}
\mathcal L_w(\widehat Q_{A,w,c})
-
\inf_{Q\in\mathcal Q_A}\mathcal L_w(Q)
\leq
\varepsilon_c,
\qquad
\varepsilon_c\longrightarrow0.
\label{eq:computeoptimization}
\end{equation}
The left-hand side is the remaining optimisation gap at computational budget $c$. The limit $\varepsilon_c\to0$ is not asserted as a universal convergence theorem for neural optimisation; it isolates the case in which additional computation solves the same fixed likelihood problem increasingly well, without by itself changing the target distribution, hypothesis class or inductive principle. Holding $P_w$, $\mathcal Q_A$ and the objective fixed, reducing $\varepsilon_c$ approaches the same optimum more accurately. It does not introduce a Solomonoff mixture over program s. If additional computation is instead used to enlarge the model class, then $\mathcal Q_A$ changes, but that is an architectural expansion. It still does not follow from the likelihood objective alone that the enlarged family is ordered according to algorithmic probability.

The stronger recent claim that loss minimisation itself should be interpreted as program-length optimisation therefore requires assumptions beyond the cross-entropy identity~\cite{wanmei2025}. The decomposition in \eqref{eq:curatedloss} fixes the population optimum from the supplied distribution; it contains no explicit $2^{-|p|}$ weighting and no program enumeration. A trained network may nevertheless internalise compression, algorithms or a bounded approximation to universal prediction. The distinction here is between what a particular trained system may implement and what its likelihood objective mathematically entails.

Wan and Mei's construction is usefully read as complementary to this distinction rather than as a contradiction of it. Starting from a trained LLM, they combine the model with an arithmetic encoding of the observed string, decoding information, a random seed and a prefix-code construction on a universal machine. This defines a computable family of programs and a corresponding surrogate mass, which may be written schematically as
\begin{equation}
\overline M_{\theta}(x)
=
\sum_s 2^{-\ell(\overline f_{\theta}(x,s))}
\leq
M(x).
\label{eq:wanmeisurrogate}
\end{equation}
The surrogate $\overline M_{\theta}$ sums the algorithmic weights of the explicit program family constructed from the trained model; the inequality records that this computable family contributes only part of the universal program mass $M(x)$, which also includes all other valid generating program s. Wan and Mei then relate reduction of language-model loss to shortening the arithmetic description of $x$, thereby increasing the algorithmic mass contributed by this particular computable program family~\cite{wanmei2025}. Nothing in the cross-entropy calculation above excludes such a construction. A likelihood-trained model can be used as a compression component inside a larger algorithmic program family, and improving that model can improve the resulting surrogate.

The refinement concerns what follows from this construction. The inequality $\overline M_{\theta}(x)\leq M(x)$ establishes that the constructed programs contribute legitimate mass within a Solomonoff-style universal mixture, but an increasing lower bound is not by itself a convergence statement for the full universal semimeasure. In particular, reducing the code length generated by one model-dependent program family does not alone establish that the remaining Solomonoff mass from all other programs becomes negligible. Wan and Mei's later formulation makes the corresponding condition explicit at inference time: their surrogate next-token interpretation is stated under the assumption that $M(x_{1:t})\approx\overline M(x_{1:t})$~\cite{wanmei2026}. Under such an assumption, the program-based surrogate can yield a normalised continuation distribution closely related to the LLM's next-token probabilities. The substantive algorithmic step is therefore the construction of $\overline M$, together with the assumption that it adequately approximates $M$, rather than the cross-entropy identity alone.

The two results consequently operate at different levels. Equation~\eqref{eq:curatedloss} is an objective-level statement: for a supplied target distribution, unrestricted population log-loss is minimised by matching that distribution. Wan and Mei provide a construction-level statement: a trained likelihood model can be wrapped in explicit coding and universal-machine machinery so that its code lengths define a computable subset of Solomonoff-style program mass. Read this way, their result refines rather than contradicts the present argument. It supplies a concrete example of the additional algorithmic structure through which a likelihood-trained LLM may participate in a Solomonoff-style estimator, while the present analysis identifies why that conclusion is not entailed by the entropy-based objective in isolation. This distinction also clarifies the role of CTM below: CTM makes the program-space approximation explicit by estimating algorithmic probability from enumerated computational mechanisms rather than deriving the program  family only from the code lengths of a particular trained predictor.

\subsection{Why conventional compression is a weak test of algorithmic induction}

A further methodological issue concerns the use of conventional compression
algorithms, such as LZW, or entropy-optimised neural predictors as evidence
for a ``simplicity bias'' related to Algorithmic Information Theory. Such
comparisons can be informative about statistical redundancy, but they should
not be interpreted as strong evidence for approximation to algorithmic
probability or Solomonoff induction.

LZW is a lossless dictionary-based compressor. Its success depends on
discovering repeated substrings and exploiting regularities that reduce
empirical description length. In this sense, it certainly exhibits a form of
simplicity preference: sequences with repeated or statistically regular
structure are compressed more effectively than sequences lacking such
redundancy. However, this is not the same object as algorithmic probability.

For a reference universal prefix machine $U$, Solomonoff--Levin algorithmic
probability is defined schematically by

$$m_U(x)=\sum_{p:U(p)=x}2^{-|p|},$$

where the probability assigned to $x$ arises from the aggregate weight of all
programs $p$ on $U$ that generate it, with each program weighted exponentially
by its description length. Solomonoff induction therefore organises prediction
over a universal generative hypothesis space. LZW does not enumerate such
programs, does not weight hypotheses by universal program length, and does not
operate over a universal mixture of computable generators.

The distinction is therefore stronger than the observation that both
compression and algorithmic probability reward ``simplicity''. A conventional
compressor can prefer statistically simple strings without approximating
Kolmogorov complexity in the algorithmic sense. Repetition, low entropy rate,
short-period structure, local dependencies, or regularity within a restricted
grammar can all make an object easy to compress for LZW while saying little
about the shortest program or family of programs capable of generating the
object.

In particular, the Shannon entropy rate of a stationary source,

$$h=\lim_{n\to\infty}\frac{1}{n}H(X_1,\ldots,X_n),$$

characterises asymptotic statistical compressibility under suitable conditions.
A compressor that asymptotically approaches this rate demonstrates successful
exploitation of statistical redundancy. But this fact alone does not imply
approximation to $K_U(x)$ or to $m_U(x)$. Algorithmic probability contains
ordinary statistical compressibility as a special and comparatively weak case:
a regular source may be statistically compressible because of repetition or
predictable frequencies, whereas the distinctive content of AIT concerns the
length and distribution of generative programs capable of producing the data.

This creates a potential circularity when an entropy-based or dictionary-based
compressor is used to validate another entropy-optimised predictor. If one
method is sensitive to statistical redundancy and another is evaluated by how
well it reproduces the same redundancy, agreement may largely establish that
both systems detect compressible statistical structure. In that sense,
``compression indicates simplicity'' and ``simplicity predicts compression''
can become two formulations of the same entropic fact. Such an experiment may
verify a necessary condition for Solomonoff-style induction, because any
sufficiently general predictor should exploit statistically visible regularity,
but the condition is too weak to identify the specifically algorithmic content
of Solomonoff induction.

The same issue applies to neural predictors trained with likelihood or
cross-entropy objectives. Their loss can be written as

$$H(P,Q)=H(P)+D_{KL}(P\Vert Q),$$

so optimisation rewards agreement with the supplied predictive distribution.
A neural model may learn rich internal structure and may even implement
algorithmic procedures, but low cross-entropy or strong predictive compression
does not by itself establish that the model is assigning mass according to
$2^{-|p|}$ over a universal program space. Predictive success is therefore
compatible with, but not diagnostic of, Solomonoff-like induction.

The distinction becomes sharper when stated in terms of computational grammar.
Choosing a black-box compressor does not remove the choice of computational
formalism; it merely makes that choice implicit. A particular compressor,
finite-state predictor, or neural architecture operates within the effective
grammar and hypothesis class permitted by its implementation. If that computational system is not universal in the relevant sense, it is closer to
selecting a particular restricted Turing machine than to selecting a reference
Universal Turing Machine. Opacity does not confer universality.

This is important because replacing an explicit reference UTM by an arbitrary
black-box predictor is sometimes presented as avoiding the arbitrariness of
reference-machine choice. It does not. The black box still embodies a specific
computational language, architecture, representation and inductive bias. The
difference is that these restrictions are less explicit. A non-universal
predictor may therefore be more restrictive than an explicitly chosen UTM,
not more general.

By contrast, a UTM-based construction has a principled theoretical status.
Although algorithmic complexity depends on the choice of reference UTM up to
the usual additive constant, all universal machines share the ability to
represent the full class of computable functions. For universal machines $U$
and $V$, the invariance theorem gives

$$K_U(x)\leq K_V(x)+c_{UV},$$

for some machine-dependent constant $c_{UV}$ independent of $x$. The practical
importance of the constant can be substantial for short objects, but this is
fundamentally different from restricting inference to a computational class
that is not universal at all.

CTM-style methods make this distinction explicit. Rather than using
compression performance as an indirect proxy for simplicity, they approximate
algorithmic probability by enumerating or systematically sampling computational
mechanisms within a specified universal formalism and measuring their output
frequencies. In a finite approximation with program-length and runtime bounds
$L$ and $T$, one may write

$$m_{U,L,T}(x)=\sum_{\substack{|p|\leq L\\U(p)=x\ \text{within}\ T}}
2^{-|p|}.$$

As the accessible program and runtime classes are enlarged, the approximation
ranges over an increasingly rich subset of the program space of the reference
UTM. Because exact algorithmic probability is lower semicomputable rather than
computable, this does not provide an effective finite procedure for obtaining
the exact universal semimeasure. It does, however, provide a principled limiting
connection to a well-defined algorithmic object that an arbitrary statistical
compressor does not acquire merely through improved predictive performance.

Moreover, the reference-machine issue should not be caricatured as the arbitrary
selection of a single UTM. Previous CTM work has examined multiple universal
computational formalisms and machine classes and has studied empirical stability,
convergence and robustness of resulting complexity rankings across reference
choices. The relevant comparison is therefore not between an ``arbitrary UTM''
and a supposedly assumption-free black box. It is between an explicitly
specified universal program space, whose machine dependence can be studied,
and an implicit computational grammar whose restrictions may be harder to
characterise and which may not be universal.

For these reasons, compressors such as LZW, entropy-rate estimators, and
likelihood-trained neural predictors are useful statistical controls but weak
tests of Solomonoff-like induction. Stronger evidence requires experiments in
which statistical compressibility is controlled while program-level generative
structure varies. Ideally, one would compare sequences with similar entropy
rate, local redundancy, or conventional compression ratio but different
algorithmic generative complexity, and then ask whether the predictor follows
the program-induced conditional distribution rather than merely the statistical
surface.

This distinction is also supported by the empirical results in Section~7.
There, the CTW baseline achieved substantially better next-bit coding than the
Transformer while failing to improve out-of-support extrapolation or exact
mechanism recovery. Better predictive compression therefore did not imply
better access to the generating mechanism. This separates ordinary coding
performance from explicit mechanism-space inference and illustrates why
entropic compression, while necessary for broad predictive competence, is too
weak by itself to establish approximation to algorithmic probability or
Solomonoff induction.

Returning to the broader data--computation trade-off, the distinction can now be located more precisely.

If a system learns to select, reweight or generate its own training data and validates those transformations against independent evidence, that curation process becomes part of the inference architecture. In that case it would be misleading to describe the complete system as merely a cross-entropy estimator. This point strengthens rather than weakens the architectural thesis: likelihood objectives can participate in a Solomonoff-style or mechanism-seeking system, but the additional search, curation and validation machinery must be included when explaining where the inductive power comes from.

The Coding Theorem Method (CTM) approaches that finite regime by enumerating large classes of small programs and estimating algorithmic probability from their output frequencies~\cite{delahayezenil2012}. BDM extends the strategy to larger objects by decomposing them into smaller components for which CTM estimates can be reused~\cite{zenil2018}. Algorithmic Information Dynamics uses related estimates under controlled perturbations to study candidate causal contribution, dynamical structure and reprogram mability~\cite{zenil2019,zenil2023}. None of these approximations is a causal oracle. Their relevance here is that they search a space organised by generative description rather than only by similarity in an observed sample.

The three algorithmic tools play different roles in this argument and should not be conflated. CTM is the most direct connection to algorithmic probability. By exhaustively or systematically running small computational devices, it estimates the frequency with which finite outputs are generated and thereby supplies empirical approximations to algorithmic probability for objects small enough to be covered by the chosen machine space. Its value is not that the approximation is universal in practice, but that it provides a non-Shannon estimate sensitive to candidate generative structure.

BDM addresses the immediate scaling problem. Exact CTM tables necessarily cover only relatively small objects. BDM decomposes a larger object into parts for which algorithmic-complexity estimates are available and combines those estimates while accounting for repeated components. This introduces decomposition dependence and cannot reproduce the full Kolmogorov complexity of an arbitrary large system, but it provides a practical bridge between local algorithmic estimates and larger structured data. In the present paper BDM is enabling rather than foundational: it shows how the mechanism-oriented bias need not remain confined to toy strings.

Once a candidate object or mechanism has been represented, the question for self-improvement is no longer only what compact process could generate this object but also which components matter to its behaviour, and what changes when they are perturbed or intervened? AID uses changes in estimated algorithmic information under controlled perturbations as one way to study structural contribution and reprogram mability. This is closer to the intervention side of the self-improvement problem. A mechanism that can be represented but not meaningfully modified is of limited use to an architecture expected to redesign itself or its models.

The distinction is important because none of these methods supplies truth automatically. CTM can privilege a short explanation that happens to be wrong. BDM can obscure dependencies that cross a chosen decomposition. AID can assign significance differently under different representations or perturbation schemes. Their role is therefore to propose and organise structural hypotheses that can be executed or perturbed. Epistemic work is completed only when the consequences of these hypotheses are compared with independent evidence or constraints.

The difference can be illustrated with a familiar toy example. From the finite sequence $2,4,6,8$, a program such as ``generate successive even numbers'' implies $10,12,\ldots$ even though those values were not observed. The inference is not guaranteed to be true: infinitely many other programs agree with four observations and diverge later. The example nevertheless captures the function of model synthesis. A generative hypothesis can entail unobserved consequences. Those consequences then become predictions to be checked. A purely empirical replacement distribution containing only the observed values has no comparable entailment unless additional structure has been learned or supplied.

The finite computational character of this search is central. If program length and runtime are bounded, define
\begin{equation}
m_{L,T}(x)=
\sum_{\substack{|p|\leq L:\ U(p)=x\ \mathrm{within}\ T\ \mathrm{steps}}}2^{-|p|}.
\label{eq:boundedmass}
\end{equation}
This finite version of algorithmic probability counts only programs no longer than $L$ that halt within $T$ steps. Increasing either bound can admit candidate generators that were previously computationally inaccessible and therefore cannot reduce the finite mass. A candidate mechanism excluded by an earlier computational budget can become accessible later, while the observations remain unchanged. This is exactly the resource-bounded situation discussed in the previous section. No information is created; the search has made a different part of the hypothesis space operationally available.

The same construction can be expressed as a bounded next-token predictor. Let $M^{\mathrm{pre}}_{L,T}(x)$ denote the algorithmic mass of programs of length at most $L$ whose outputs begin with the prefix $x$ within $T$ steps,
\begin{equation}
M^{\mathrm{pre}}_{L,T}(x)
=
\sum_{\substack{|p|\leq L:\ U(p)\text{ outputs a string with prefix }x\\
\text{within }T\text{ steps}}}2^{-|p|}.
\label{eq:boundedprefixmass}
\end{equation}
The programs need not halt exactly after producing $x$; it is enough that their outputs begin with the observed prefix. The quantity therefore measures finite algorithmic support for continuations compatible with the current context. For a next token $a\in\mathcal V$, this induces the finite continuation distribution
\begin{equation}
\pi_{L,T}(a\mid x)
=
\frac{M^{\mathrm{pre}}_{L,T}(xa)}
{\sum_{b\in\mathcal V}M^{\mathrm{pre}}_{L,T}(xb)},
\label{eq:boundedsolomonoffpredictor}
\end{equation}
whenever the denominator is positive. The normalisation compares the program-weighted mass supporting each possible continuation $xa$ across the vocabulary. Unlike the learned softmax in \eqref{eq:softmaxnext}, the scores in \eqref{eq:boundedsolomonoffpredictor} are explicitly determined by the generating programs admitted by the bounded search. This is still not exact Solomonoff induction: the program  length, runtime and machine class are finite, and the universal semimeasure remains uncomputable. It nevertheless makes explicit what a resource-bounded Solomonoff-style next-token estimator would have to add beyond likelihood-based token prediction.

A CTM-style enumeration supplies finite empirical approximations to algorithmic probability in precisely this generative direction. An open-ended implementation can progressively enlarge the enumerated machine or program  classes and runtime budgets, thereby expanding the set of generative explanations that are operationally available rather than merely refining parameters inside a fixed token-level objective. BDM can make related algorithmic estimates usable on larger structured objects for which direct enumeration is infeasible, while AID can use perturbations of candidate structures to identify components whose modification changes estimated algorithmic organisation or behaviour. BDM and AID do not themselves turn the bounded predictor into the exact Solomonoff semimeasure, nor do they guarantee monotonic improvement in predictive accuracy. Their role is to make the finite algorithmic search more scalable and intervention-oriented. In that qualified sense, an open-ended CTM process extended by BDM and AID provides a route towards progressively richer, resource-bounded Solomonoff-style induction rather than a claim to compute Solomonoff induction exactly.

The difference in the use of computation can also be written directly. Let $c$ index a computational resource budget and, for an observed prefix $x$, define the accessible generating program s
\begin{equation}
\mathcal G_c(x)
=
\left\{p:\ |p|\leq L(c),\ U(p)\text{ outputs a string with prefix }x
\text{ within }T(c)\text{ steps}\right\},
\label{eq:computebudgetprogram s}
\end{equation}
where $L(c)$ and $T(c)$ are non-decreasing program-length and runtime limits. The set $\mathcal G_c(x)$ is thus the program hypothesis space that can actually be searched at budget $c$. Since $c_1<c_2$ implies $\mathcal G_{c_1}(x)\subseteq\mathcal G_{c_2}(x)$, additional computation can make new generative mechanisms available rather than merely optimise parameters within a fixed hypothesis class. The bounded prefix mass in \eqref{eq:boundedprefixmass} is precisely a weighted sum over such an accessible set. Increasing computation in this construction can make a previously excluded generative mechanism enter the hypothesis space. This is categorically different from merely reducing the optimisation error $\varepsilon_c$ in \eqref{eq:computeoptimization} while the likelihood target and hypothesis class remain fixed.

Seen in this way, CTM changes the role assigned to additional computational power. In likelihood-based training, more computation devoted to the same objective principally improves optimisation or supports a larger chosen model, but does not by itself supply the universal program prior. In an open-ended CTM regime, increased computation can enlarge the program classes and runtimes from which algorithmic probability is estimated. BDM provides a route for applying related estimates to structures beyond the scale of direct enumeration, while AID provides a perturbational framework for examining how candidate structures respond to modification. The resulting system therefore places progressively more of the search for explanatory structure inside the computational process itself. This does not remove dependence on the reference machine, representation or empirical evidence, but it gives increased computation a qualitatively different inferential role.

The elementary even-number example is intentionally simple, but the same distinction appears whenever observations underdetermine a mechanism. Consider two dynamical systems that agree over the observed time window but diverge after a perturbation. A flexible predictor may fit both histories equally well. A mechanism-oriented learner instead attempts to construct candidate update rules and can ask which intervention separates them. The advantage is not guaranteed correctness. It is that the hypothesis is expressed in a form from which discriminating consequences can be derived.

A second example is scientific extrapolation outside observed support. Suppose a small collection of measurements is consistent with both a low-degree law and a high-capacity interpolating function. Within the measured range their predictive errors may be indistinguishable. The compact law makes a stronger commitment about what should happen elsewhere. That commitment is risky, but precisely because it is risky it can be tested. Mechanism synthesis converts fit into a source of counterfactual or extrapolative questions. A distributional model can of course learn the same law internally; the point is functional, not architectural. What matters is whether the system can expose and manipulate the structure sufficiently to derive and test such commitments.

This is the sense in which model synthesis is closer to scientific inference than unrestricted generation. A scientific model earns epistemic value not because it produces novel statements, but because it compresses observations into a structure that entails further consequences and survives attempts to falsify them. The structure may be an equation, a causal graph, a program , a simulator or another executable representation. Symbolic model synthesis is used in this paper as a collective name for that function, not as a claim that classical symbolic AI is the only implementation capable of it.

A neurosymbolic architecture is useful here if the term is understood functionally. Statistical models are powerful proposal generators and estimators. Symbolic objects such as program s, equations, causal graphs, type constraints, conservation laws and proof objects provide forms that can be executed, composed, perturbed or checked. An effective system need not choose between them. A neural component can propose a candidate mechanism, an algorithmic or symbolic component can make its structure explicit, and an external observation, intervention or verifier can decide whether the consequences survive testing.

The central benefit is not that symbolic structure magically restores lost data or guarantees a contraction toward truth. Structural priors can be wrong. A short program may fit the observations for the wrong reason. A causal graph may be unidentifiable from the available interventions. CTM estimates depend on the finite machine class and runtime cutoff, BDM on decomposition, and AID on the perturbation model and interpretation. Model synthesis therefore needs independent validation rather than replacing it.

Four practical failure modes are worth distinguishing. There is a representation failure when the chosen language contains no concise adequate mechanism; a budget failure when an adequate mechanism exists but lies outside the current length or runtime limits; an identification failure when several accessible mechanisms remain compatible with the evidence; and a search or verification failure when an accessible mechanism is not found or is tested inadequately. These are operationally different problems. Calling them all ``lack of information'' hides precisely the methodological distinctions that matter in empirical discovery.

The functional loop proposed here is consequently simple to state. Generate candidate mechanisms, derive consequences, test those consequences against evidence or invariant rules, and retain the validated changes. CTM, BDM and AID are possible ingredients in the mechanism-proposal and structural-analysis stages, not exceptions to information conservation. Their role is to enlarge or reorganise access to mechanism space. The direction of improvement still comes from successful contact with a criterion that is not merely an echo of the current model distribution.

There is an older epistemological distinction behind this argument. Novel expression is not the same as new warranted knowledge. Generative models make this unusually visible because they can produce strings, images, proofs and hypotheses that have never previously occurred while remaining entirely within a fixed computational interaction with their inputs and parameters. Novelty in the output space is therefore cheap evidence for epistemic progress. The harder question is whether the generated object establishes a relation to the world, a formal system or an engineering constraint that was not previously warranted.

The Kantian analytic/synthetic distinction offers an analogy, although not a machine-learning theorem. An inference can unpack consequences of what is already represented without thereby acquiring a new empirical fact. Scientific progress typically combines such internal derivation with observations that can defeat the current theory. Algorithmic search occupies the first side of this cycle when it exposes consequences implicit in data and priors; experiments and interventions occupy the second when they alter the evidential state. The strength of a self-improving architecture lies in coupling them.

This perspective also helps distinguish explanation from prediction. Predictive adequacy is indispensable, but a mechanism has additional uses when it supports intervention, counterfactual reasoning, composition with other mechanisms or transfer to regimes not represented in the training distribution. None of these properties follows merely from assigning high probability to observed data. Conversely, an explicit mechanism that predicts badly is not rescued by being interpretable or simple. The argument is for a cycle in which generative structure makes stronger commitments and empirical or formal evaluation decides whether those commitments deserve to survive.

\section{Finite empirical tests of resource-bounded mechanistic search}
\label{sec:empirical}

\subsection{Empirical protocol}

For illustrative purposes, we tested the finite operational predictions of
Eqs.~\eqref{eq:boundedmass}--\eqref{eq:computebudgetprogram s} in an explicit
prefix-free program space rather than attempting to approximate exact
Solomonoff induction. The benchmark contained $11{,}250$ unique deterministic
binary mechanisms generated by four families---periodic rules, affine modular
threshold rules, XOR-lag recurrences, and Walsh-like generators. Forty-eight
ambiguous target mechanisms were selected, balanced equally across the four
families ($12$ per family). Each target generated a $128$-bit sequence. For
the fixed-observation experiment, only the first $16$ bits were observed and
held unchanged while the accessible description-length budget was increased
from $L=4$ to $26$ bits; prediction was evaluated on the following $64$ bits.

For each accessible budget, the finite program predictor assigned a mechanism
the complete bounded mass
\[
m_L(x)=\sum_{\substack{p:U(p)=x\\ |p|\leq L}}2^{-|p|},
\]
summing over all program descriptions in the finite language rather than using
only the shortest description. We compared this predictor with a uniform
mixture over prefix-compatible mechanisms, a likelihood-only order-$8$
Krichevsky--Trofimov autoregressor, a depth-$8$ Context-Tree Weighting
(CTW-D8) predictor prespecified to match the order-$8$ statistical baseline,
and a small task-local Transformer trained
solely by next-token cross-entropy. The Transformer used two encoder layers,
four attention heads, embedding dimension $48$, AdamW
($\mathrm{lr}=3\times10^{-3}$, weight decay $10^{-4}$), and $450$ optimisation
steps on the same observed prefix. It is intended as a controlled
likelihood-based neural baseline, not as a proxy for a frontier LLM.

Two additional structural priors were computed independently over the same
candidate mechanisms. BDM was evaluated with 
\texttt{pybdm} implementation in one dimension A second codelength prior used the bundled
\texttt{pycdm} implementation and its symmetrised certified codelength. These
quantities are finite complexity estimators. CTW-D8 was included as a stronger Bayesian/universal-coding statistical
baseline: it performs context-tree mixture prediction without explicit search
over the finite mechanism class used by the program-space predictors.

A second experiment compared three recursive update loops from the same initial
eight observed bits over $14$ update cycles. In closed self-training, each
additional accepted bit was generated by the currently selected mechanism
itself. In the random-verifier condition, a randomly selected unresolved
position was provided by the true external mechanism. In the active
program-plus-verifier condition, the unresolved position of maximal posterior
Bernoulli entropy among the highest-mass candidate mechanisms was queried and
its true externally supplied value was added to the evidence. Outcomes were
external $64$-bit continuation accuracy, consistency with accepted evidence,
exact mechanism recovery, and restricted mean cycles to recovery.

All primary comparisons were paired by target mechanism. Means and differences
are reported with $4{,}000$-resample bootstrap $95\%$ confidence intervals.
Paired comparisons use $10{,}000$ sign-permutation draws with
Benjamini--Hochberg false-discovery-rate correction, and exact-recovery
proportions use Wilson intervals.

\subsection{Resource expansion exposes predictive mechanisms under fixed evidence}

\begin{figure}[htbp]
\small
\centering
\makebox[\textwidth][c]{%
\includegraphics[width=1\textwidth]{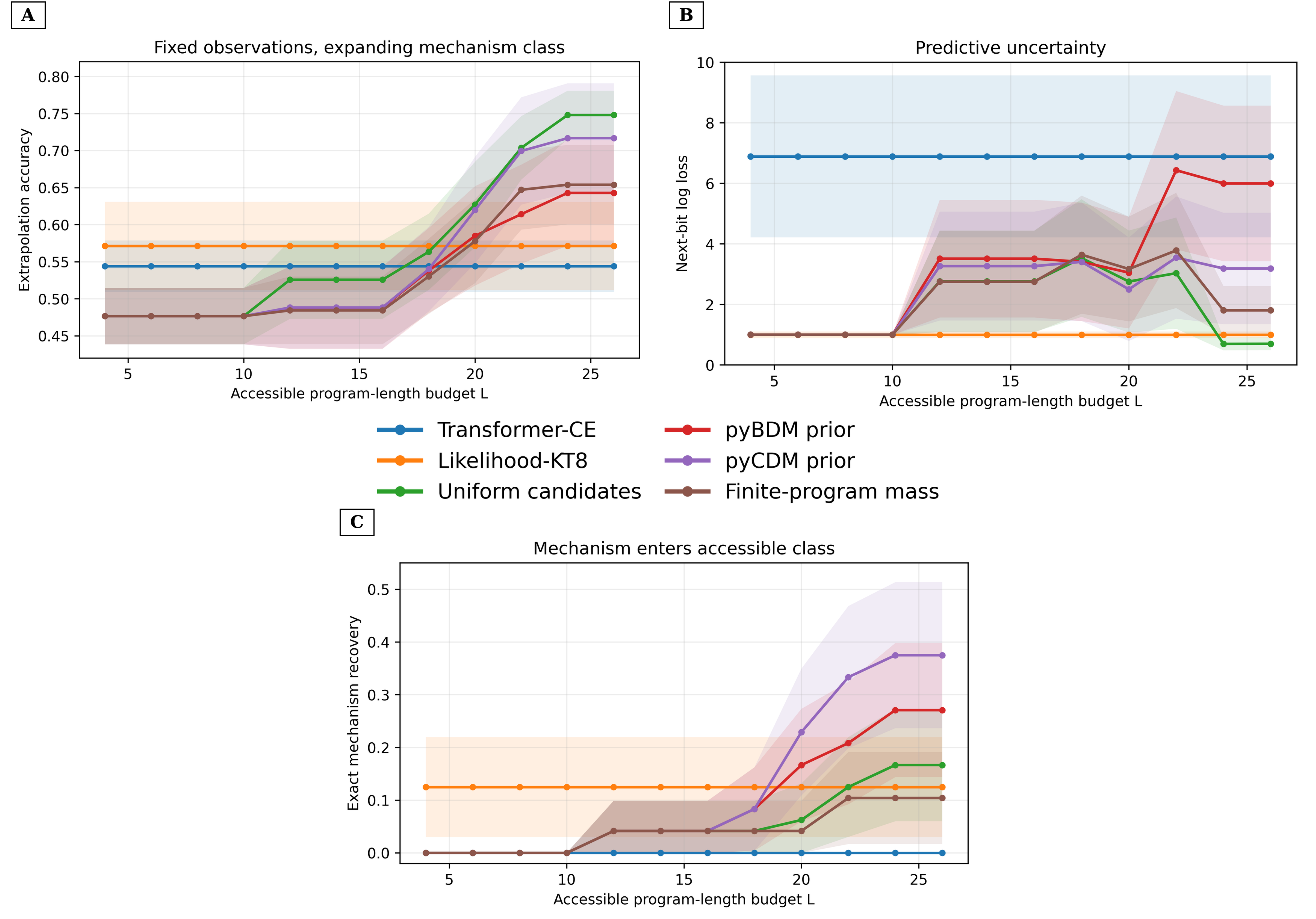}
}
\caption{\textbf{Fixed observations with an expanding accessible mechanism class.} Forty-eight ambiguous short-program environments were evaluated while the first $16$ observations remained unchanged and the maximum admissible program length $L$ increased. \textbf{(A) Held-out $64$-bit extrapolation accuracy.} With the observed prefix fixed, increasing the accessible program-length budget $L$ improves extrapolative accuracy only for predictors that search an explicit compatible mechanism class. Likelihood-only baselines remain flat. Shading denotes $95\%$ bootstrap confidence intervals across target mechanisms. \textbf{(B) Next-bit log loss.} Access to additional compatible mechanisms changes predictive uncertainty for the mechanism-space methods, whereas the fixed likelihood baselines do not improve with larger search budgets. Lower values indicate better local predictive coding. Shading denotes $95\%$ bootstrap confidence intervals across target mechanisms. \textbf{(C) Exact recovery of the generating continuation.} Enlarging the accessible mechanism class increases the fraction of targets for which the correct generator enters the supported search space and is recovered exactly, again in contrast to likelihood-only baselines. Shading denotes $95\%$ bootstrap confidence intervals across target mechanisms. The three panels separate extrapolative accuracy, local predictive coding, and exact mechanism recovery. The experiment tests finite resource-bounded accessibility and does not claim computation of exact Solomonoff induction.}
\label{fig:resourceempirical}
\end{figure}

Holding the observed prefix fixed while enlarging the accessible mechanism
class produced a clear gain in extrapolative performance
(Fig.~\ref{fig:resourceempirical}a). At the maximum program budget, the
cross-entropy Transformer achieved mean continuation accuracy
$0.544$ (95\% CI $0.512$--$0.579$). The uniform mechanism mixture reached
$0.748$ ($0.717$--$0.780$; paired $\Delta=0.204$, FDR
$q=5.0\times10^{-4}$), the pyCDM prior reached $0.717$
($0.644$--$0.792$; $\Delta=0.173$, $q=5.0\times10^{-4}$), the finite
program-mass predictor reached $0.654$ ($0.600$--$0.705$;
$\Delta=0.110$, $q=8.3\times10^{-4}$), and the pyBDM prior reached
$0.643$ ($0.574$--$0.714$; $\Delta=0.099$, $q=0.0215$).
The likelihood-only KT baseline reached $0.571$ and did not significantly
differ from the Transformer after correction ($q=0.4168$). CTW-D8 likewise
showed no extrapolation advantage, reaching $0.538$
(95\% CI $0.493$--$0.585$; $\Delta=-0.006$, FDR $q=0.7523$), with exact
mechanism recovery of only $2.1\%$. As Fig.~\ref{fig:resourceempirical}b
shows, CTW nonetheless achieved substantially lower next-bit log loss
($0.814$ [0.601, 1.041]) than the Transformer ($6.890$ [4.230, 9.635])
without improving out-of-support extrapolation, separating local predictive
coding from explicit mechanism-space inference as argued herein.

Exact mechanistic recovery further confirmed the same distinction
(Fig.~\ref{fig:resourceempirical}c): the Transformer recovered none of the
$48$ target mechanisms, whereas pyCDM recovered $37.5\%$ (95\% Wilson CI
$25.2$--$51.6\%$), pyBDM $27.1\%$, the uniform mechanism mixture $16.7\%$,
and finite program mass $10.4\%$. The result supports the
resource-accessibility claim rather than a universal ordering of complexity
priors: the uniform candidate mixture gave the strongest average
extrapolation and next-bit calibration, whereas pyCDM gave the highest exact
mechanism recovery. What consistently separated the successful methods from
the fixed likelihood baseline was access to an explicit compatible mechanism
class whose contents changed with the search budget.

\begin{table}[p]
\small
\centering
\normalsize
\renewcommand{\arraystretch}{1.40}
\setlength{\tabcolsep}{3pt}

\caption{\textbf{Fixed-observation prediction at the maximum accessible
program budget.} Confidence intervals are bootstrap $95\%$ intervals for
accuracy and log loss and Wilson intervals for exact recovery. $q$ values are
Benjamini--Hochberg-adjusted paired comparisons with the Transformer.
CTW-D8 provides a stronger statistical/universal-coding baseline: despite
competitive next-bit coding, it does not reproduce the extrapolation or
mechanism-recovery gains of explicit mechanism-space search. Across the
mechanism-space methods, uniform candidate averaging gives the strongest mean
extrapolation, whereas pyCDM gives the highest exact mechanism recovery.}
\label{tab:resourceempirical}

\begin{tabular}{
>{\raggedright\arraybackslash}p{0.15\textwidth}
>{\centering\arraybackslash}p{0.15\textwidth}
>{\centering\arraybackslash}p{0.17\textwidth}
>{\centering\arraybackslash}p{0.15\textwidth}
>{\centering\arraybackslash}p{0.15\textwidth}
>{\centering\arraybackslash}p{0.08\textwidth}
}

\toprule

\textbf{Method} &
\makecell[c]{\textbf{Extrapolation}\\\textbf{accuracy}} &
\makecell[c]{\textbf{$\Delta$ accuracy}\\\textbf{vs.\ Transformer}} &
\makecell[c]{\textbf{Next-bit}\\\textbf{log loss}} &
\makecell[c]{\textbf{Exact}\\\textbf{recovery}} &
\makecell[c]{\textbf{FDR}\\\textbf{$q$}} \\

\midrule

Transformer-CE &
\makecell[c]{0.544 [0.511,\\0.579]} &
-- &
\makecell[c]{6.890 [4.230,\\9.635]} &
\makecell[c]{0.000 [0.000,\\0.074]} &
-- \\

\addlinespace[4pt]

Likelihood-KT8 &
\makecell[c]{0.571 [0.514,\\0.632]} &
\makecell[c]{0.027 [-0.025,\\0.085]} &
\makecell[c]{0.998 [0.895,\\1.104]} &
\makecell[c]{0.125 [0.059,\\0.247]} &
0.4168 \\

\addlinespace[4pt]

CTW-D8 &
\makecell[c]{0.538 [0.493,\\0.585]} &
\makecell[c]{-0.006 [-0.038,\\0.028]} &
\makecell[c]{0.814 [0.601,\\1.041]} &
\makecell[c]{0.021 [0.004,\\0.109]} &
0.7523 \\

\addlinespace[4pt]

Uniform candidates &
\makecell[c]{0.748 [0.716,\\0.781]} &
\makecell[c]{0.204 [0.162,\\0.248]} &
\makecell[c]{0.700 [0.508,\\0.917]} &
\makecell[c]{0.167 [0.087,\\0.296]} &
0.0006 \\

\addlinespace[4pt]

pyBDM prior &
\makecell[c]{0.643 [0.573,\\0.710]} &
\makecell[c]{0.099 [0.024,\\0.177]} &
\makecell[c]{5.997 [3.505,\\8.605]} &
\makecell[c]{0.271 [0.166,\\0.410]} &
0.0266 \\

\addlinespace[4pt]

pyCDM prior &
\makecell[c]{0.717 [0.645,\\0.788]} &
\makecell[c]{0.173 [0.095,\\0.249]} &
\makecell[c]{3.187 [1.474,\\5.105]} &
\makecell[c]{0.375 [0.252,\\0.516]} &
0.0008 \\

\addlinespace[4pt]

Finite-program mass &
\makecell[c]{0.654 [0.600,\\0.706]} &
\makecell[c]{0.110 [0.054,\\0.167]} &
\makecell[c]{1.805 [1.044,\\2.619]} &
\makecell[c]{0.104 [0.045,\\0.222]} &
0.0008 \\

\bottomrule

\end{tabular}
\end{table}

\subsection{Independent verification converts recursion into cumulative improvement}

\begin{figure}[H]
\centering
\makebox[\textwidth][c]{%
\includegraphics[width=1\textwidth]{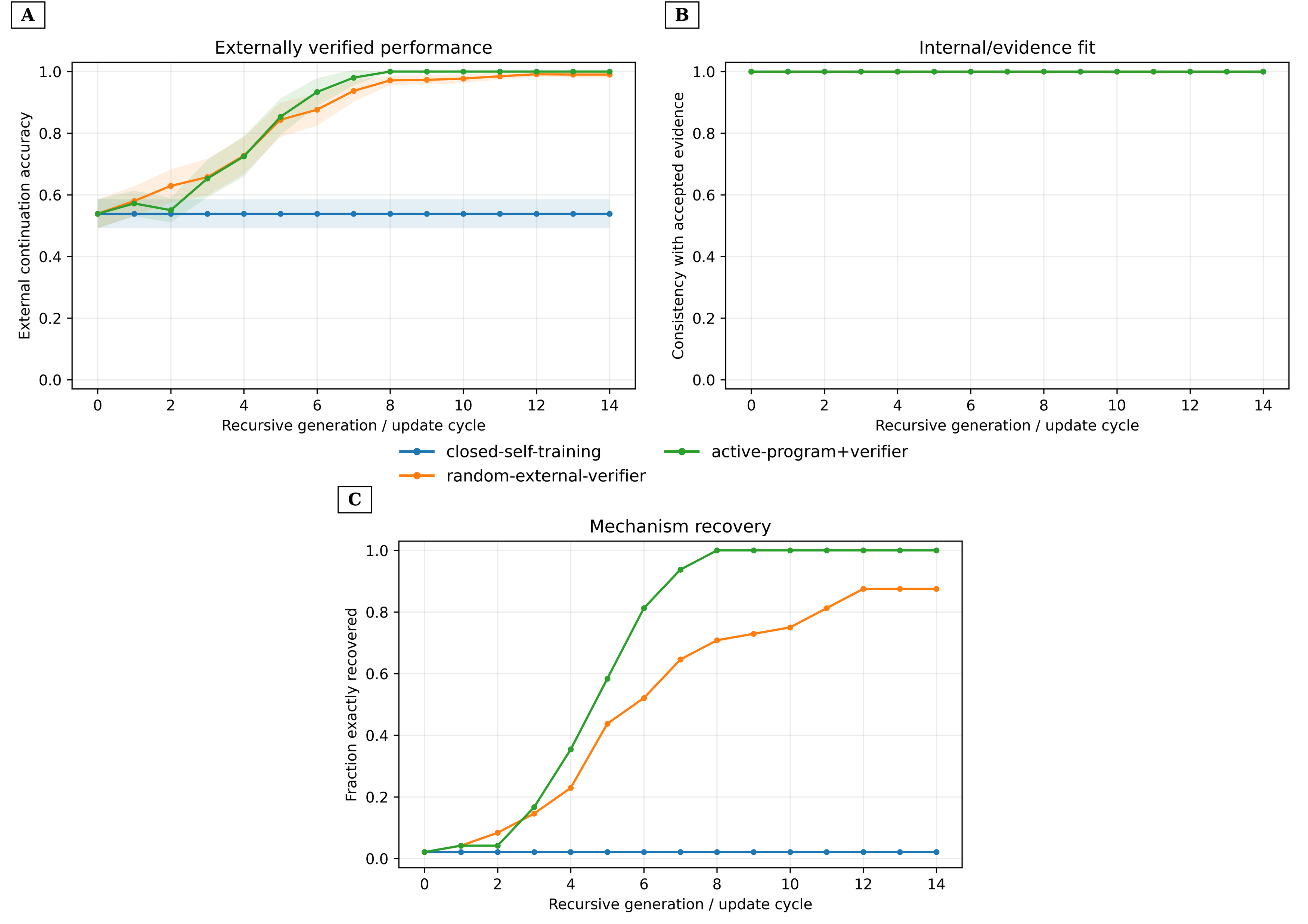}
}
\caption{\textbf{Recursive self-training versus independently verified mechanism search.} \textbf{(A) Externally evaluated continuation accuracy.} Closed self-training stays nearly flat, while independently verified recursion rapidly improves external performance. Random verification helps, and active program-guided querying helps most. Shading denotes $95\%$ bootstrap confidence intervals across target mechanisms. \textbf{(B) Consistency with accepted evidence.} All three loops remain perfectly consistent with the evidence they accept, showing that internal agreement alone does not diagnose actual improvement. \textbf{(C) Exact recovery of the generating mechanism.} Independent verification progressively restores the true mechanism, and active program-guided querying achieves complete recovery earlier than random verification. The three panels disentangle external task improvement, internal evidence fit, and exact recovery of the hidden generator over recursive update cycles. Closed self-training remains internally coherent without improving externally, whereas independently verified recursion restores the true mechanism.}
\label{fig:recursiveempirical}
\end{figure}

The recursive experiment directly separated internal consistency from externally
verified improvement. Figure~\ref{fig:recursiveempirical}a shows that closed
self-training remained nearly flat in external performance, while
Fig.~\ref{fig:recursiveempirical}b shows that it nevertheless remained
perfectly consistent with the evidence it had accepted. Its final external
continuation accuracy remained $0.538$ (95\% CI $0.494$--$0.587$),
essentially its initial level, and exact mechanism recovery in
Fig.~\ref{fig:recursiveempirical}c was only $2.1\%$. Thus recursive agreement
with self-generated evidence was not itself an improvement signal.

Replacing endogenous evidence by independent observations changed the
dynamics. Random external verification produced final external accuracy
$0.990$ ($0.979$--$0.998$), exact recovery $87.5\%$, and a mean restricted
time to recovery of $7.25$ cycles. Active mechanism-guided querying followed
by the same independent verifier reached external accuracy $1.000$, exact
recovery $100\%$, and restricted mean time to recovery of $5.04$ cycles.
Relative to closed self-training, the external-accuracy gains were
$0.452$ and $0.462$, respectively, with FDR-adjusted
$q=9.999\times10^{-5}$ for both comparisons. Crucially,
Fig.~\ref{fig:recursiveempirical}b shows that consistency with the accepted
evidence was $1.0$ in all three conditions throughout, whereas
Fig.~\ref{fig:recursiveempirical}c shows that only the independently verified
loops progressively recovered the correct generating mechanism. Internal
agreement therefore failed to distinguish the stationary closed loop from the
two externally improving loops, whereas independent verification did.

\begin{table}[ht]
\small
\centering
\normalsize
\renewcommand{\arraystretch}{1.45}
\setlength{\tabcolsep}{4pt}

\caption{\textbf{Recursive-loop outcomes after $14$ update cycles.}
Restricted mean cycles count unrecovered targets at the fixed
$15$-cycle censoring boundary. Accuracy comparisons are paired against
closed self-training and FDR corrected.}
\label{tab:recursiveempirical}

\begin{tabular}{
>{\raggedright\arraybackslash}p{0.17\textwidth}
>{\centering\arraybackslash}p{0.16\textwidth}
>{\centering\arraybackslash}p{0.14\textwidth}
>{\centering\arraybackslash}p{0.18\textwidth}
>{\centering\arraybackslash}p{0.25\textwidth}
}

\toprule

\textbf{Loop} &
\makecell[c]{\textbf{Final external}\\\textbf{accuracy}} &
\makecell[c]{\textbf{Exact}\\\textbf{recovery}} &
\makecell[c]{\textbf{Restricted mean}\\\textbf{cycles to recovery}} &
\makecell[c]{\textbf{$\Delta$ Accuracy vs.}\\\textbf{\ closed FDR $q$}} \\

\midrule

\makecell[l]{Closed\\self-training} &
\makecell[c]{0.538 [0.494,\\0.587]} &
\makecell[c]{0.021 [0.004,\\0.109]} &
\makecell[c]{14.69 [14.06,\\15.00]} &
-- \\

\addlinespace[6pt]

\makecell[l]{Random external\\verifier} &
\makecell[c]{0.990 [0.979,\\0.998]} &
\makecell[c]{0.875 [0.753,\\0.941]} &
\makecell[c]{7.25 [6.15, 8.44]} &
\makecell[c]{0.452 [0.399, 0.500]\\
$q=9.999\times10^{-5}$} \\

\addlinespace[6pt]

\makecell[l]{Active program +\\verifier} &
\makecell[c]{1.000 [1.000,\\1.000]} &
\makecell[c]{1.000 [0.926,\\1.000]} &
\makecell[c]{5.04 [4.56, 5.50]} &
\makecell[c]{0.462 [0.414, 0.505]\\
$q=9.999\times10^{-5}$} \\

\bottomrule

\end{tabular}
\end{table}

\subsection{Direct measurements on served large language models}
\label{sec:servedllm}

\begin{figure}[!htbp]
\centering
\includegraphics[width=\textwidth]{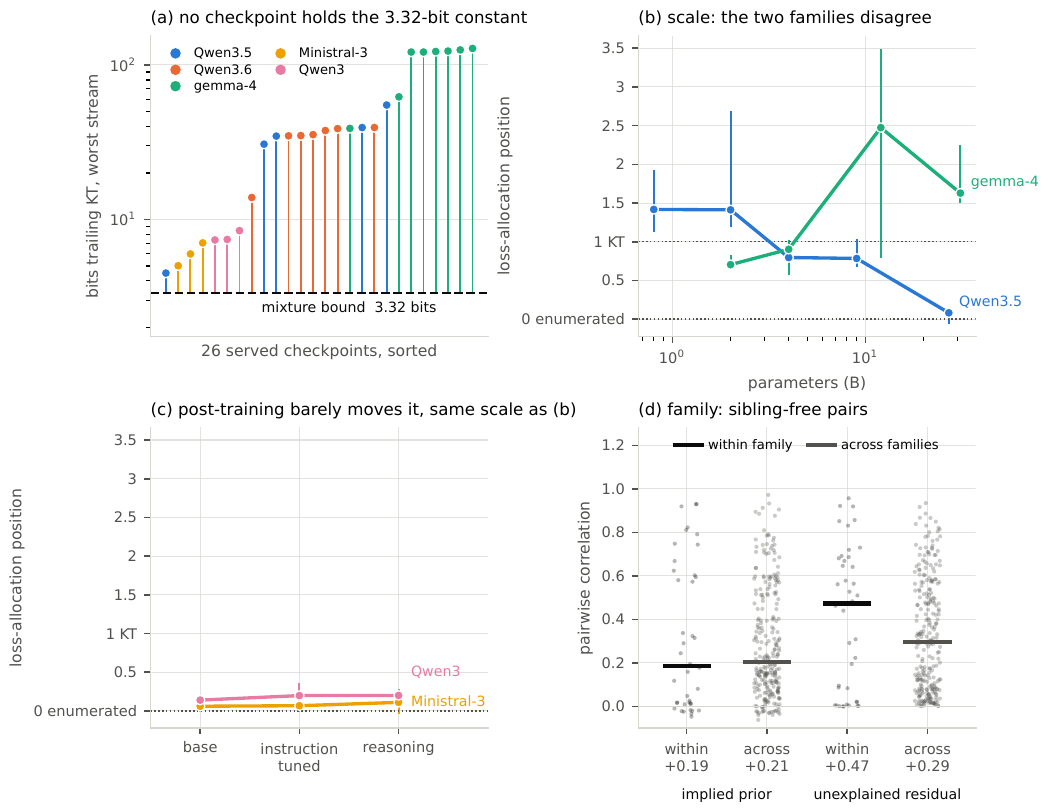}
\caption{\textbf{Direct measurements on twenty-six served language-model checkpoints.}
\textbf{(a)} Dominance is the guarantee that makes a mixture worth its cost: bounded regret against every member of
the class. A mixture carrying KT among its components at weight $0.10$ trails it by at most $\log(1/0.10)=3.32$ bits
on any stream (dashed line), a bound the reference system of this paper attains at $3.316$ bits. Every checkpoint
tested lies above it at twenty-four symbols; colour is the pre-training family. \textbf{(b)} Where each model's loss
allocation sits between an enumerated mixture over the class ($0$) and the complexity-blind KT anchor ($1$), against
parameter count, for the two families measured at eight bits. The two families move in opposite directions, but the
disagreement is confounded with architecture class (gemma's two smallest rungs are E-series checkpoints), the position
tracks off-alphabet mass outside one family, and it changes with the rendering; the wide interval at $12$B reflects
that deployment's large serving floor. \textbf{(c)} The same quantity across post-training stages, on the same
vertical scale as (b): holding pre-training and size fixed, post-training moves it little, in two families that both
sit near the enumerated end. \textbf{(d)} Pairwise correlation between distinct checkpoints (rungs of one precision
ladder excluded) of the implied prior, and of the residual direction that no mixture over the class explains, split
into within-family and across-family pairs (points jittered; bars are medians). The implied priors show no family
effect; the residual directions are more alike within families, an exploratory finding. Whiskers in (b) and (c) span
that checkpoint's independent designs.}
\label{fig:llmevidence}
\end{figure}

The experiments above use an explicit program space and a deliberately small
task-local Transformer. That design isolates the mechanism-search question, but
it leaves the paper's title claim, which is about large language models,
supported only by analysis. This section reports direct measurements on served models, taken from the companion
study~\cite{ozelimzenil2026b}, which gives the full protocol, its validation and the
analyses behind every number. It is written to be read on its own: the class, the rendering, the estimator and the
reference values are all stated here, and the numbers quoted are those of
Table~\ref{tab:llmevidence} and Figure~\ref{fig:llmevidence}.

\paragraph{The declared class.} Asking whether a predictor is a program-weighted
mixture only has an exact answer if the mixture can actually be computed, so the
tests are posed over an enumerated class rather than over all programs. The class
contains $177{,}591$ deterministic generators of sequences over the three-symbol
alphabet $\{0,1,2\}$: the three constants, every periodic pattern of period two
to five, and every order-one and order-two rule together with its initial
condition. Each generator $h$ is given a description length $\ell(h)$ in bits,
the length of the code that names it, and a prior weight proportional to
$2^{-\ell(h)}$. Short descriptions therefore receive more weight, which is the
feature of Solomonoff's prior that the class is meant to reproduce in computable
form.

The description length is not a free parameter; it is read off the program.
Naming one symbol of a three-letter alphabet costs $\log_2 3 = 1.585$ bits, so a
program pays that much for each symbol it has to specify, plus a codeword
identifying which family it belongs to, plus $\log_2 L$ bits for a periodic
program to state its period $L$. The family codewords are two bits for the
constants and the periodic programs and three bits for the two rule families,
and that asymmetry is forced rather than chosen. Because each program pays
$\log_2 3$ per named symbol, the number of programs in a family cancels their
individual weights exactly, so a family's total contribution to the Kraft sum
depends only on the length of its codeword: the constants contribute $1/4$, the
period-$L$ programs $1/(4L)$, and each rule family $1/8$, for a total of
$0.8208$. Had the two rule families also been given two-bit codewords they would
have contributed $1/4$ each and the sum would be $1.0708$, exceeding one, so no
prefix-free code with those lengths exists and $2^{-\ell(h)}$ could not be
normalised into a prior at all. The third bit is what buys the headroom.
Three examples make the arithmetic explicit. A constant names one symbol and
pays $\log_2 3 + 2 = 3.58$ bits. The period-two pattern $01$ names two symbols
and its period, paying $2\log_2 3 + 2 + \log_2 2 = 6.17$ bits. An order-one rule
must name the three entries of its transition table $f$ and one initial symbol,
so it pays $(3+1)\log_2 3 + 3 = 9.34$ bits; an order-two rule names nine table
entries and two initial symbols, paying $(9+2)\log_2 3 + 3 = 20.43$ bits.
Table~\ref{tab:class} lists every family with its count, and writes four programs
out in full together with the sequence each emits.

The prior weight follows mechanically. A program of length $\ell(h)$ receives
$2^{-\ell(h)}$, and these are divided by $Z=\sum_h 2^{-\ell(h)}=0.8208$ so that
they sum to one. The constant therefore receives $2^{-3.58}/0.8208=0.102$, the
largest weight in the class, and the order-two rule receives
$2^{-20.43}/0.8208=8.6\times10^{-7}$. The simplest program carries roughly
$117{,}000$ times the weight of the most complex one. That ratio is the whole
content of the prior: it is what makes short descriptions preferred, and it is
exactly what a complexity-blind code such as KT does not have.

The class is numerically dominated by its most expensive family, since $177{,}147$
of the $177{,}591$ programs are order-two rules, but it is not dominated by them
in weight, each carrying under a millionth of the mass. At the horizon of $40$
symbols used here many programs emit identical sequences, so the $177{,}591$
hypotheses collapse to $3{,}945$ distinct \emph{behavioural classes}. Those
classes, not the programs, are the coordinates in which two predictors' implied
priors are compared, because a predictor can only be held responsible for
behaviour it could in principle distinguish.

\begin{table}[t]
\centering
\small
\caption{\textbf{The declared class, and how a program's description length is computed.} Every length is built from three parts: $\log_2 3 = 1.585$ bits for each symbol the program must name, a family codeword of $2$ bits for the constants and periodic programs and $3$ bits for the two rule families, and, for a periodic program, $\log_2 L$ bits to state its period $L$. The third bit on the two rule families is forced, not chosen: with two-bit codewords throughout the Kraft sum would be $1.0708>1$, so no prefix-free code with those lengths would exist. \emph{Upper:} the seven families and their counts. \emph{Lower:} four programs written out in full, with the sequence each emits, its length, and the prior weight $2^{-\ell(h)}/Z$ that length earns it, where $Z=\sum_h 2^{-\ell(h)}=0.8208$. The simplest program carries about $117{,}000$ times the weight of the most complex, which is the preference for short descriptions that the class exists to supply and that a complexity-blind code does not have.}
\label{tab:class}
\begin{tabular}{lrlr}
\toprule
\textbf{Family} & \textbf{Programs} & \textbf{Description length} & \textbf{bits} \\
\midrule
constant & $3$ & $\log_2 3 + 2$ & $3.58$ \\
periodic{,} period $2$ & $9$ & $2\log_2 3 + 2 + \log_2 2$ & $6.17$ \\
periodic{,} period $3$ & $27$ & $3\log_2 3 + 2 + \log_2 3$ & $8.34$ \\
periodic{,} period $4$ & $81$ & $4\log_2 3 + 2 + \log_2 4$ & $10.34$ \\
periodic{,} period $5$ & $243$ & $5\log_2 3 + 2 + \log_2 5$ & $12.25$ \\
order-one rule & $81$ & $(3+1)\log_2 3 + 3$ & $9.34$ \\
order-two rule & $177{,}147$ & $(9+2)\log_2 3 + 3$ & $20.43$ \\
\midrule
\textbf{total} & $\mathbf{177{,}591}$ & & \\
\bottomrule
\end{tabular}

\bigskip

\resizebox{\textwidth}{!}{%
\begin{tabular}{llrr}
\toprule
\textbf{Program} & \textbf{Emits} & \textbf{$\ell(h)$ (bits)} & \textbf{Prior weight} \\
\midrule
output $1$ for ever & \texttt{11111111111111\ldots} & $\log_2 3 + 2$ $= 3.58$ & $1.02 \times 10^{-1}$ \\
repeat the block $01$ & \texttt{01010101010101\ldots} & $2\log_2 3 + 2 + \log_2 2$ $= 6.17$ & $1.69 \times 10^{-2}$ \\
$x_{t+1}=f(x_t)$, $f=[1,2,0]$, $x_0=0$ & \texttt{01201201201201\ldots} & $(3+1)\log_2 3 + 3$ $= 9.34$ & $1.88 \times 10^{-3}$ \\
$x_{t+1}=f(x_t,x_{t-1})$, $x_0x_1=01$ & \texttt{01101101101101\ldots} & $(9+2)\log_2 3 + 3$ $= 20.43$ & $8.60 \times 10^{-7}$ \\
\bottomrule
\end{tabular}}
\end{table}

\paragraph{The two reference values.} Two predictors over this class bound the
comparison and give the measurements their meaning.

The \emph{enumerated mixture} is the Bayes mixture $M(x)=\sum_h w_h P_h(x)$ with
the weights above. It is the closest computable analogue of a Solomonoff
predictor: it weights hypotheses by description length and it inherits the
property that matters here. Because $M(x)\geq w_c P_c(x)$ for every component
$c$, the mixture's cumulative log-loss can exceed that of any component by at
most $\log(1/w_c)$ bits, whatever the data. That is \emph{dominance}, and it is
the reason a universal mixture is worth its cost: it cannot be badly beaten by
anything it contains, ever, on any sequence.

The \emph{Krichevsky--Trofimov} (KT) code is a standard universal code for
memoryless sources. It is complexity-blind: it assigns no weight to a
hypothesis' description length, so it cannot prefer a simple generator to a
complicated one. A Solomonoff-style predictor should behave like the mixture and
unlike KT, and the gap between them is what the measurements below locate.

Putting the two together gives a falsifiable number. A mixture that includes KT
among its components with weight $w$ trails KT by at most $\log(1/w)$ bits on
any stream. In a reference system built from this class, with KT carried at
weight $0.10$, the bound is $\log(1/0.10)=3.32$ bits, and the system measures
$3.316$ bits: the bound is tight and it is attained. A neural predictor trained
by next-token cross-entropy on samples from the same class, scored on the same
streams, instead trails KT by $112.4$ bits and is still losing ground at $5.1$
bits per step. The same quantity therefore separates a predictor that has the
guarantee from one that does not, on the same class and the same data, before
any language model is involved. It is this quantity that is measured on served
models below.

\paragraph{Reducing a language model to a predictor on that class.} A served
model is made comparable to these references by reading it as a next-symbol
predictor on the class's streams. Symbols are rendered as bare digits after a start-of-document token, one token
per symbol; the rendering is verified per checkpoint, and tokenisers that group
digits are excluded rather than silently rendered differently. A grammar
restricted to the three digits is applied at the server as a mask. A read
requests the full-vocabulary probability table of the generated position before
the sampler chain, so the returned values are a plain softmax of the raw logits
and are unaffected by temperature, by the mask or by a logit bias; the three
digits' probabilities are then renormalised. The tests therefore concern a
well-defined object: the masked, per-step renormalised predictor. The
renormalisation discards the mass the model puts on other tokens, up to $83\%$
for one checkpoint, and what is discarded depends on the rendering; this matters
for the allocation readings below.

\paragraph{The serving floor, and why every number is referred to it.} A served
model is not a fixed conditional. Repeated reads of the same context under
batched serving return slightly different probabilities, so the deployment is a
random measure $Q_w$ and the estimand is its mean $\bar Q = \mathbb E_w Q_w$.
Reading each context several times gives an unbiased estimate of the replicate
variance, the \emph{serving floor}, whose reference value is zero for a
deterministic server. The floor bounds what any of these tests can resolve: a
departure smaller than the floor is not evidence of agreement, only of
insufficient resolution. Every quantity below is reported against it.

\paragraph{The panel.} Twenty-six served checkpoints span five pre-training
families, $0.8$ to $35$ billion parameters, dense and sparse mixture-of-experts
architectures, serving precisions from $1.9$ to $8.5$ bits per weight, and base,
instruction-tuned and reasoning post-training stages. They fill twenty-seven
places in the design, because the four-bit Qwen3.6-35B-A3B checkpoint serves both
as a rung of its precision ladder and as the sparse half of the architecture
pair; it is one checkpoint, collected once and read once, entered under two
contrasts. Each checkpoint is read on three to five independent randomised
designs, whose probe sets are drawn by seed and do not depend on the model,
giving $88$ distinct collections and ninety-one checkpoint--designs once that
shared entry is counted in both of its arms. The collections comprise
$1{,}339{,}755$ reads and $2{,}121{,}524$ restricted trajectories. No trajectory left the alphabet in any checkpoint or design. Results by contrast are in
Table~\ref{tab:llmevidence}.

\begin{table}[t]
\centering
\small
\setlength{\tabcolsep}{4pt}
\caption{\textbf{Served-checkpoint measurements by contrast.} Each cell is the median over the checkpoints of that contrast, with the range across them where they differ; each checkpoint is itself the median over its three to five independent designs. Column~2 is the number of bits by which the served model trails a Krichevsky--Trofimov code on its worst stream at twenty-four symbols: a mixture that carries KT among its components at weight $0.10$ trails it by at most $3.32$ bits, and no checkpoint stays within that. Column~3 counts checkpoint--designs in which the hypothesis that the served conditionals are \emph{some} mixture over the declared class is rejected under a per-checkpoint Holm correction; over a class of deterministic generators this is a test of support. Column~4 locates the loss allocation between an enumerated mixture over the class ($0$) and the complexity-blind KT anchor ($1$); outside one family it tracks the mass a checkpoint puts outside the alphabet, and it changes with the rendering. Column~5 is the serving floor, the replicate spread of the served conditional, which bounds what any of these tests can resolve.}
\label{tab:llmevidence}
\resizebox{\textwidth}{!}{%
\begin{tabular}{lcccc}
\toprule
\textbf{Contrast (what varies)} & \textbf{Bits behind KT} & \textbf{Membership rejected} & \textbf{Allocation} & \textbf{Floor} \\
\midrule
Scale, Qwen3.5 ($0.8$--$27$B, \texttt{Q8\_0}) & 35 [4, 55] & 21/25 & 0.80 [0.08, 1.42] & 0.015 [0.004, 0.030] \\
Scale, gemma-4 ($2$--$31$B, \texttt{Q8\_0}) & 92 [39, 125] & 9/12 & 1.27 [0.70, 2.47] & 0.014 [0.008, 0.240] \\
Precision, Qwen3.6-35B-A3B ($1.9$--$8.5$ bpw) & 36 [35, 39] & 18/18 & 0.81 [0.73, 0.89] & 0.124 [0.113, 0.145] \\
Precision, gemma-4-31B ($2.6$--$8.5$ bpw) & 123 [121, 128] & 11/15 & 1.40 [1.31, 1.63] & 0.034 [0.018, 0.043] \\
Architecture, Qwen3.6 \texttt{Q4\_K\_M} (dense vs.\ sparse) & 26 [14, 39] & 6/6 & 0.51 [0.29, 0.74] & 0.067 [0.010, 0.125] \\
Post-training, Ministral-3-14B & 6 [5, 7] & 9/9 & 0.07 [0.06, 0.11] & 0.004 [0.003, 0.004] \\
Post-training, Qwen3-30B-A3B & 7 [7, 8] & 9/9 & 0.20 [0.14, 0.20] & 0.008 [0.007, 0.011] \\
Family at comparable size & 92 [62, 122] & 3/6 & 1.69 [0.90, 2.47] & 0.124 [0.008, 0.240] \\
\midrule
\textbf{Whole panel} ($26$ checkpoints) & 36 [4, 128] & 80/91 & 0.80 [0.06, 2.47] & 0.022 [0.003, 0.240] \\
\bottomrule
\end{tabular}}
\end{table}

\paragraph{No checkpoint holds the reference mixture's constant.}
Measured against the $3.32$-bit reference above, all twenty-six served checkpoints
exceed it on their worst stream at twenty-four symbols: the best, a $27$B dense model,
trails KT by $4.5$ bits, the median checkpoint by $36.5$ bits, and the worst by $128$
bits (Figure~\ref{fig:llmevidence}a). The $3.32$ bits is the constant of a mixture that
gives KT weight $0.10$; a bare model states no constant, and could still dominate KT
with a larger one.

A bound is a constant, so the decisive question is whether the excess settles, and it
needs a longer horizon than twenty-four symbols, where even the reference mixture is
still approaching its bound. Read to $256$ symbols, the answer differs by checkpoint.
The three nearest the bound settle: from $96$ to $256$ symbols their worst-stream
excess stays within $6.7$--$8.3$ bits (the $27$B Qwen3.5), $7.2$--$7.7$ bits (the
$30$B Qwen3 base) and $12.0$--$13.1$ bits (the $14$B Ministral base). Their worst case
is a uniform random stream, on which they trail KT at a roughly fixed distance, as a
mixture carrying KT at a weight between about $2^{-7}$ and $2^{-13}$ would. Three read
further from the bound do not settle: their excess keeps growing, as roughly $n^{0.26}$
to $n^{0.59}$. No served model therefore holds the reference mixture's constant; those
that hold any constant over the horizon read hold one two to four times larger, and
the others hold none. The three that settle are also those that put the most mass
outside the alphabet, so this reading carries the confound described below.

This is the sharpest finite-horizon evidence we have for the position argued in
Section~\ref{sec:solomonoff}: on this class, likelihood training does not deliver the
constant that a mixture carrying a complexity-blind code at weight $0.10$ holds, and in
some checkpoints it delivers no constant at all.

\paragraph{Served conditionals are not a mixture over the declared class.} A
weaker question than dominance is whether the served conditionals are consistent
with \emph{some} mixture over the declared enumerable class, under any weights
rather than the correct ones. Solving $q=Gw$ with $w\geq0$ on shared probes and referring the
residual to the serving floor, the hypothesis is rejected, under a
per-checkpoint Holm correction, in $80$ of the $91$ checkpoint--designs. The
eleven exceptions are deployments whose floors are too large to resolve a
departure, not deployments that pass: nine have a residual at or below their own
floor and two sit just above it ($1.07$ and $1.09$ times) without surviving the
correction, while the $80$ rejecting designs lie $1.2$ to $114$ times above their
floor, with a median of $5$. Read one request at a time, where the floor collapses
to zero, every checkpoint rejects. Over a class of deterministic generators this is a
test of support: every checkpoint puts mass on continuations that no member of the
class produces, which any model not confined to the class must. On this class and this
rendering, then, the served conditionals are not a mixture over the declared class.

Section~\ref{sec:solomonoff} argues that additional computation spent on a fixed
likelihood objective need not change the inductive principle. The panel gives a
sharper result than a null. Under a description-length prior the description
length of the generating hypothesis bounds the achievable loss rather than
predicting its mean, so the quantity to estimate is the upper envelope: we take
the slope of a line through the per-stratum $0.9$ quantiles of realised loss
against description length, and rescale it so that the enumerated mixture reads
$0$ and the complexity-blind KT anchor reads $1$. A predictor that allocates
loss by description length in the way the class's prior does sits near $0$; one
that ignores description length sits near $1$. On that scale,
the two families measured move in \emph{opposite directions}
(Figure~\ref{fig:llmevidence}b). One falls from $1.42$ at $0.8$B to $0.08$ at $27$B;
the other rises from $0.70$ at $2$B to $1.63$ at $31$B across four rungs at the same
quantisation. Three things keep this from being read as an effect of scale. The rise
crosses an architecture boundary, since gemma's two smallest rungs are E-series
checkpoints. Outside that family the position tracks, almost rank for rank, the mass a
checkpoint puts outside the alphabet, which the renormalisation discards. And re-read
with a space after each digit, the first family's fall from $0.8$B to $9$B disappears
($0.76$, $0.92$, $0.71$) and only the step to $27$B remains, while the $12$B gemma moves
from $2.47$ to $-1.74$. No effect of scale on allocation is therefore established, in
either direction; a single-family study would have reported a clean trend.

\subsection{Post-training does not supply a new inductive principle}

The most
common objection to the present argument is that instruction tuning or
reasoning-style post-training introduces the missing structure. Two arms of the
panel hold pre-training and size fixed and vary only the post-training stage. A
$14$B checkpoint as base, instruction-tuned and reasoning model gives slope
positions $0.06$, $0.07$ and $0.11$; a $30$B mixture-of-experts as base,
instruction-tuned and thinking model gives $0.14$, $0.20$ and $0.20$
(Figure~\ref{fig:llmevidence}c, drawn on the same scale as the scale arm). In both
families the movement is far smaller than the gap between any two pre-training
families. Post-training changes what the model emits; on these rows it leaves the
allocation where pre-training put it. Three qualifications bound that reading: both
families sit near the enumerated end, where the position has little room to move;
their base checkpoints come from a different conversion pipeline, which could mask a
real effect as easily as create one; and the result holds under one rendering.

The
closed-loop concern of Section~\ref{sec:recursive} is that agreement between a
generator and its evaluator can grow while competence does not. The panel makes
that measurable, because the probes do not depend on the model and every implied
prior lives in the same coordinates. Rungs of one precision ladder are near-duplicates
of one checkpoint, so pairs of them are set aside. Across the remaining pairs the
implied priors show no family effect: they correlate $+0.19$ within a pre-training
family and $+0.21$ across ($p = 0.61$ by permutation). The component that no mixture
over the class explains, the fitted residual direction, is more alike within families,
$+0.47$ against $+0.29$ ($p = 0.014$; Figure~\ref{fig:llmevidence}d); that test was
chosen after the first, so it is exploratory, and part of the correlation may be
forced by the geometry of the fit on a shared probe set. Read with those caveats,
models that disagree about the class depart from it in a partly common direction,
which is consistent with the closed-loop concern: an evaluator drawn from the same
population as the generator may not be independent of the generator's characteristic
error.

\subsection{A finite observer has a measured ceiling.} Section~\ref{sec:conservation}
argues that formal information-conservation results set a ceiling that does not
by itself choose an empirical method, and that finite resources decide what is
operationally accessible. The serving floor is a concrete instance of exactly
that. Across the panel it spans $0.003$ to $0.240$, a factor of eighty, and it
is predicted neither by weight precision (it is flat across a four-fold range of
bit-width within a single checkpoint) nor by architecture class (one
mixture-of-experts family sits among the lowest floors, while the largest floor
belongs to a dense model), and it depends on the rendering: the largest, $0.240$,
falls to $0.055$ when the same checkpoint reads digits separated by spaces. Below its
floor a deployment's conditionals cannot be resolved, however capable the underlying
model. The ceiling here is imposed by the serving stack and the interface rather than
by the weights, and it must be measured per deployment and rendering rather than
inferred from the specification.

Every number in this section and in
Table~\ref{tab:llmevidence} is produced by a seeded script from cached server
responses, so the analysis can be re-run offline without the models. The panel
definition, covering checkpoint, file, repository revision, quantisation and
serving flags, together with the collection driver, the battery and the table and figure generators
are released with the paper; the raw cached reads and trajectories
are the only inputs the analyses use, and each checkpoint--design is a separate
cache.

The rendering specified above was chosen because it makes the reference values
exact, and it is a deliberately favourable setting for the models: the alphabet is
closed, the class is enumerable, and no restricted trajectory left it. The claims
above are therefore claims about served language models read as next-digit predictors
on this class under this rendering, not about language models in general, and they
do not show that a neural system cannot implement algorithmic induction. What they
establish is narrower and sufficient for the present argument: the systems currently
described as approximating Solomonoff induction do not hold the dominance constant of
a mixture that carries a complexity-blind code at weight $0.10$; some hold a larger
one over the horizon read and others none; and their allocation cannot yet be
separated from the interface through which it is read.

\section{Consequences for AGI, ASI and the so-called Singularity}

The results in this paper isolate a missing step in a common narrative. The ability to generate one's own future training data does not establish an improvement operator. Exact self-training is stationary; finite self-replacement may drift or lose diversity; and a corrective signal changes the process precisely because it introduces a criterion not contained in neutral self-imitation.

Good's intelligence explosion requires a process in which improvements to the system increase its capacity to produce further improvements~\cite{good1966}. Vinge and Kurzweil place related feedback at the centre of their accounts of technological discontinuity~\cite{vinge1993,kurzweil2005}. For contemporary generative AI, the question is therefore concrete: which part of the system proposes a change, which part decides that the change is genuinely better, what evidence supports that judgement, and how does the system prevent the evaluator from becoming circular as the generator changes?

A closed recursive language-modelling loop has an incomplete answer. It can produce vast numbers of candidates, including code, plans and critiques. It can also imitate evaluation language. But if candidate generation and evaluation are ultimately trained to agree with the same endogenous distribution, agreement is not an independent measure of truth, causal adequacy or general capability. More generations of the same loop do not solve that epistemic problem simply by increasing scale.

Externally anchored systems are different. A model that writes faster code and runs it against a trusted performance benchmark has a real improvement signal. A model that proposes a proof and submits it to a sound proof checker has a correctness criterion independent of its confidence. A model that designs an experiment and receives a measurement from the world changes its information state. A model that plays a formal game can improve through self-play because the rules and outcome remain invariant even when all trajectories are synthetic. None of these systems is closed in the relevant sense.

The distinction also prevents a misleading equation between autonomy and collapse. A future system could be highly autonomous while continuously seeking external evidence. Indeed, greater capability may allow it to design better experiments, build better instruments or find stronger formal tests. Such a system would not resemble a passive stream of fresh human-authored text, yet it would remain epistemically open. The concern addressed here applies when the feedback loop becomes progressively self-referential, not when human supervision merely decreases.

This is where symbolic model synthesis enters the picture. An intelligence explosion, if it is to involve open-ended scientific and technological competence rather than only increasingly efficient distribution fitting, needs a way to represent candidate mechanisms in forms that can be modified and tested. program synthesis is one route. Causal models, equations, formal proofs and executable simulators are others. The requirement is functional: the system must be able to move from a predictive surface to candidate generative structure and back to independent tests.

Nothing in this argument proves that only CTM or AID can provide that capability. Nor does it establish that a transformer cannot internalise the relevant operations. If a neural system reliably constructs executable mechanisms, derives counterfactual or extrapolative consequences and tests them against independent evidence, then it has implemented the required model-synthesis layer, whatever its internal architecture is called.

The next-token formulation sharpens this architectural distinction. Two systems can both return a probability distribution over the next token while implementing very different inductive procedures. A conventional LLM obtains that distribution from learned logits optimised primarily for likelihood on observed token sequences. A CTM-augmented system can additionally score continuations or latent mechanisms by finite algorithmic probability, so that explanatory programs compete according to generative description rather than token fit alone. Increasing the capacity or training compute of the former can improve the learned conditional distribution without changing the defining objective. Increasing the search budget of the latter can enlarge the program  class over which algorithmic hypotheses are considered. If those hypotheses are then exposed to independent empirical or formal tests and successful mechanisms are retained, increased computational capability can improve the mechanism-search process itself. That is a plausible source of the second-order positive feedback required by recursive self-improvement, although it remains an empirical architectural hypothesis rather than a consequence of algorithmic probability alone.

The paper rejects the inference that recursive LLM generation and retraining, by themselves, already instantiate the positive-feedback mechanism assumed in classical Singularity narratives. A future system may add the missing ingredients. The claim is that those ingredients are substantive and cannot be inferred from recursive density matching alone.

The connection to Good can now be stated more sharply. An ultraintelligent machine must do more than search a fixed design space faster. If its own improvement changes the representation of the problem, the hypotheses it can formulate and the evaluators it uses, then the feedback loop must preserve a notion of success through those changes. Otherwise a system can appear to improve according to an evaluator that it has itself made easier to satisfy. The classical intelligence-explosion premise therefore contains an often implicit epistemic assumption: the process that recognises improvement remains sufficiently anchored while the system responsible for proposing improvements changes.

Vinge's argument raises the same issue from the perspective of predictability. A transition can become difficult for humans to forecast because the system explores designs or theories outside our effective search horizon. That does not remove the need for the system itself to discriminate successful from unsuccessful changes. Indeed, the farther its search moves beyond human supervision, the more the architecture depends on reliable non-circular tests. Autonomy shifts the burden of validation from people to the system's interaction with formal and empirical constraints.

Kurzweil's account emphasises accelerating returns across interacting technologies. Such acceleration can provide more computation, more data and better tools, all of which can enlarge the accessible hypothesis space. The argument here does not deny those trends. It identifies a separate requirement. More resources can accelerate search, but search becomes recursive self-improvement only when candidate changes are connected to a criterion that tracks genuine capability and when the resulting improvements feed back into the capacity to discover and validate further changes.

We argue that that the familiar inference from increasingly autonomous LLM generation to a near-term Singularity omits an architectural step. Present systems are extraordinarily capable proposal generators and increasingly capable tool users. The unresolved question is whether they possess, or are being given, a sufficiently general mechanism-synthesis and validation loop to turn recursive generation into reliable recursive improvement across changing domains.

This formulation makes the thesis falsifiable. A strong counterexample would be a genuinely distribution-only recursive system that, without functionally equivalent mechanism construction or independent evaluation, exhibits sustained externally verified discovery and improvement across changing domains. Evidence in the opposite direction would come from systems in which explicit mechanism search plus independent verification produces improvements that self-training alone cannot sustain under matched data and computational budgets.

The thesis is deliberately implementation-neutral even if we have introduced CTM, BDM and AID methods as tools closer to AIT more capable of Solomonoff induction than other more statistical tools.

A transformer that internally learns to construct executable models, calls tools to test them, revises them after failed predictions and preserves the resulting improvements would satisfy the requirement. Conversely, attaching a symbolic module that merely reformats the model's own assertions without independent evaluation would not.

This functional reading matters because neural and symbolic computation are not mutually exclusive categories at the level relevant to the argument. Neural systems can implement algorithms; symbolic systems can use statistical search; hybrid systems can move between continuous representations and discrete executable objects. The architectural distinction of interest is between a loop whose criterion is endogenous distributional agreement and a loop that can construct commitments and expose them to tests capable of rejecting the current model.

A useful way to view the scientific method in this context is as repeated alternation between compression and resistance. A theory compresses observations by replacing many facts with a smaller generative account. From that account one derives consequences. The world then resists by failing to behave as predicted, or a formal system resists by rejecting a proof or violating a constraint. The resulting discrepancy drives revision. A self-improving system does not need to imitate human scientific institutions, but an open-ended system needs some mechanism with the same logical role: proposals must encounter something they do not control.

This also explains why simply adding more real data is not the deepest solution to recursive closure. Passive data accumulation can stabilise estimation and broaden coverage, but scientific progress often depends on choosing which data to obtain. An experiment is valuable because competing models predict different outcomes. A capable agent can therefore increase its grounding not only by consuming an external stream but by actively creating informative interactions with the world. In that regime the learner's intelligence can itself improve the quality of the evidence it obtains, providing a plausible positive feedback that is absent from neutral self-training.

The converse risk is evaluator endogeneity. As more of the training and evaluation stack is generated by the system, benchmark success can cease to track the intended capability. Reward hacking is one familiar form, but the issue is broader: a system may narrow its ontology, choose experiments that cannot falsify its favoured models, or optimise formal specifications that omit the property of interest. The requirement of non-circular evaluation should therefore be understood dynamically. Evaluators, instruments and formal specifications may themselves be revised, but their revision must remain exposed to constraints that are not simultaneously rewritten to guarantee success.

These considerations suggest a research program beyond model-collapse measurements. One line of experiments should compare systems under the same fixed observations while varying the computational and representational resources available for mechanism search. Another should allow systems to select interventions and measure how efficiently they distinguish observationally equivalent hypotheses. A third should compare recursive self-training with recursive model synthesis under the same total compute, including deliberately misaligned priors and imperfect verifiers. The important outcome is not whether one method produces more novel text, but whether it generates hypotheses whose independently tested consequences improve over successive cycles.

A particularly strong test would place a learner in a family of environments whose surface statistics change while their generating mechanisms share reusable structure. Pure distribution matching should have to relearn much of the surface variation. A successful mechanism-synthesis system should increasingly exploit the reusable generative structure, propose discriminating interventions and transfer validated components to new environments. If a distribution-only recursive system achieved the same sustained, externally verified improvement without any functionally equivalent mechanism construction or independent evaluation, it would count directly against the thesis advanced here.

\section{Discussion and limitations}

The paper began from model collapse, but its main conclusion is about the structure of improvement. Collapse is one possible outcome of finite recursive replacement. Stationarity is the exact population outcome of pure self-imitation. Neither is an intelligence explosion. Once external data, replay, correction, tools, reward models, formal constraints or environmental feedback enter, the effective update operator changes and must be analysed on its own terms. This is why the empirical synthetic-data literature should not be compressed into a single universal recurrence.

The mathematics is intentionally simple. The exact diversity result uses multinomial sampling on a finite alphabet, and the grounding calculation assumes a fixed reference distribution. Real neural training adds misspecification, optimisation state, regularisation, retrieval, architecture changes, non-stationary objectives and selective data curation. These can alter the dynamics substantially. The equations isolate mechanisms rather than simulate a frontier training pipeline. They support claims about the specified loops, not universal claims about every neural system.

The same qualification applies to the comparison between likelihood-based learning and Solomonoff induction. Cross-entropy is not intrinsically tied to a particular neural architecture, and the argument does not claim that a likelihood-trained model cannot learn program-like internal structure. Nor does it claim that all useful distribution curation must be performed by humans. A system may learn to choose its own data, curricula or interventions. When it does so, those learned operations are additional components of the effective estimator and should be evaluated as such. The narrower claim is that the likelihood objective by itself neither chooses the curation function in \eqref{eq:curateddistribution} nor supplies the program weighting characteristic of Solomonoff induction. Likewise, $C_{\mathrm{mass}}$ in \eqref{eq:massintervention} measures distributional displacement only; it is not a measure of intelligence, labour or epistemic value.

The resource-bounded accessibility formulation is likewise an abstraction. To turn it into a complete theory one would need explicit complexity classes, task losses, runtime distributions and representation constraints. Its purpose here is narrower: to make clear why the truth of DPI or Levin non-growth does not settle an empirical comparison between finite procedures. Two methods can obey the same global bound while producing very different predictions because one can operationally reach a useful mechanism and the other cannot.

Our CTW control reinforces this distinction: substantially improved
next-bit universal coding did not translate into improved extrapolation or
mechanism recovery, indicating that predictive compression of the observed
prefix and explicit access to compatible generative mechanisms are
operationally distinct.

The Kantian language should also be interpreted carefully. Kant's analytic and synthetic distinction belongs to a philosophical framework that is not a theorem about machine learning~\cite{kant1998}. The useful analogy is epistemological: producing a novel string is different from establishing a new, warranted claim about the world. A system may generate highly original hypotheses from existing information, but the epistemic status of those hypotheses depends on how they are tested. 

Causality introduces its own qualifications. Observational data can sometimes identify causal structure under strong assumptions, and interventions are not the only possible source of causal information. Conversely, interventions do not automatically solve every identification problem. The paper relies only on the weaker point that observational distribution matching does not by itself select a unique causal mechanism~\cite{pearl2009}. Any causal or algorithmic model proposed by a self-improving system must still confront identifiability, misspecification and verification.

Algorithmic methods carry parallel limitations. Kolmogorov complexity is uncomputable, CTM is a finite approximation, BDM depends on decomposition choices, and AID perturbation scores require a specified representation and perturbation scheme~\cite{zenil2018,zenil2019,zenil2023}. The invariance theorem controls machine dependence only up to an additive constant, which can be important for short objects. Simplicity is a prior, not a guarantee of truth. These limitations are reasons to couple structural inference to evidence rather than reasons to abandon it.

The independent evaluator in a recursive self-improvement loop can itself fail. Benchmarks can be gamed, proof systems can formalise the wrong specification, experiments can be poorly designed, and reward models can become endogenous proxies. The architectural solution requirement is therefore not merely to add a verifier but to maintain a chain of correction in which proposed improvements remain answerable to criteria whose failure modes are themselves exposed to independent testing. As the system improves, preserving that non-circularity may be harder than generating candidate changes.

The served-model measurements of Section~\ref{sec:servedllm} are bounded in a way
worth separating from the rendering caveat stated there. They test deployed
language models as they are served, which is the object the recursive-improvement
claims are made about, but they say nothing about a system in which a likelihood
model is one component inside a larger program-based construction of the kind
discussed in Section~\ref{sec:solomonoff}. A model that fails the dominance test
when read directly may still contribute useful mass to a surrogate built around
it. The measurements therefore constrain the claim that present LLMs already
approximate Solomonoff induction; they do not constrain what a neurosymbolic
architecture built from such models might achieve, which is the possibility this
paper argues has to be built rather than assumed.

The finite experiments in Section~\ref{sec:empirical} provide some
support for these distinctions. With the observations fixed,
expanding an explicit mechanism class improved extrapolation and mechanism
recovery relative to the task-local likelihood baseline. For instance, in the recursive
experiment, perfect consistency with self-generated evidence coexisted with
stationary external performance, whereas independently verified observations
produced cumulative improvement and mechanism recovery. Further, the active
program-guided verifier reached recovery in fewer cycles than random
external querying. These results support our architectural claim that
mechanism representation and non-circular evaluation perform distinct work
that cannot be solely inferred from recursive distributional agreement. They do
not establish that any particular algorithmic prior is universally optimal:
the uniform mechanism mixture achieved the strongest mean extrapolation in the
fixed-prefix benchmark, and the pyBDM/pyCDM robustness analyses confirmed
representation dependence. Our evidence supports the
mechanism-search and epistemic-openness thesis more strongly than any claim of
universal superiority of a single complexity estimator.


The methodological thesis was tested directly by holding observations fixed while expanding the computationally accessible mechanism class. Across 48 ambiguous short-program environments, the first 16 observations remained unchanged while the admissible program-length budget increased, allowing additional compatible generators to enter the finite search space. At the maximum budget, the likelihood-trained Transformer achieved extrapolation accuracy of $0.544$ [0.511, 0.579], compared with $0.748$ [0.716, 0.781] for the uniform mechanism mixture, $0.717$ [0.645, 0.788] for pyCDM, $0.654$ [0.600, 0.706] for finite-program mass, and $0.643$ [0.573, 0.710] for pyBDM (Fig.~\ref{fig:resourceempirical}; Table~\ref{tab:resourceempirical}). To recall, the uniform mechanism mixture assigns equal prior weight to every candidate generating program consistent with the observed prefix, then averages their predictions without favouring shorter programs. Exact mechanism recovery was $0\%$ for the Transformer, compared with $37.5\%$ for pyCDM, $27.1\%$ for pyBDM, $16.7\%$ for the uniform mixture, and $10.4\%$ for finite-program mass. These results highlight the distinction in \eqref{eq:accessible}: enlarging the accessible hypothesis class can make consequences of fixed evidence computationally available without violating the information bound in \eqref{eq:dpi}.

The next-token control revealed a a dissociation between predictive coding and mechanism-level extrapolation. CTW-D8 achieved lower next-bit log loss than the Transformer, $0.814$ [0.601, 1.041] versus $6.890$ [4.230, 9.635], yet showed no corresponding extrapolation advantage: accuracy was $0.538$ [0.493, 0.585] versus $0.544$ for the Transformer (paired $\Delta=-0.006$, FDR $q=0.7523$), and exact mechanism recovery remained only $2.1\%$. The likelihood-only KT8 baseline similarly showed no significant extrapolation advantage ($q=0.4168$). Thus, substantially better next-bit coding was not accompanied by better out-of-support extrapolation or mechanism identification. By contrast, explicit mechanism space predictors improved extrapolation as compatible generators became accessible. The uniform mixture achieved the highest mean extrapolation accuracy, whereas pyCDM achieved the highest exact recovery. These results distinguish local statistical prediction from generative mechanism inference in this finite benchmark. They support access to explicit mechanism classes rather than the universal superiority of any particular algorithmic prior, and do not establish that the Transformer is incapable of implementing such inference under other architectures or training regimes.

The recursive experiment independently demonstrated a dissociation between internal consistency and externally verified improvement. All three recursive arms maintained identical perfect consistency with their accepted evidence ($1.0$) throughout, yet their external outcomes diverged sharply (Fig.~\ref{fig:recursiveempirical}; Table~\ref{tab:recursiveempirical}). Closed self-training remained approximately stationary, with final external accuracy of $0.538$ [0.494, 0.587] and only $2.1\%$ exact recovery. Random independent verification increased accuracy to $0.990$ [0.979, 0.998] and recovery to $87.5\%$, while active program-guided querying followed by the same verifier reached $1.000$ accuracy and $100\%$ recovery. Accuracy gains over closed self-training were $0.452$ and $0.462$, respectively (FDR-adjusted $q=9.999\times10^{-5}$ for both); active querying also reduced restricted mean recovery time from $7.25$ to $5.04$ cycles relative to random verification. Crucially, perfect internal agreement could not distinguish the stationary closed loop from the improving verified loops: only independent evidence progressively corrected the inferred mechanism. These finite experiments support the distinct roles of mechanism space accessibility and non-circular verification, without establishing a universal theorem about LLMs or Solomonoff induction. Reference-language sensitivity, deliberately misaligned priors, matched computational-cost comparisons, and broader replacement, accumulation, replay and correction regimes remain necessary tests of generality in prospective studies. 
Regardless, we can generally conclude that better next-bit coding did not imply better mechanism extrapolation, and perfect internal consistency did not imply recursive improvement. 


Finally, this paper does not derive a probability or date for AGI, ASI or the Singularity. It does not claim that symbolic model synthesis is sufficient for any definition of general intelligence, that other neural system architectures cannot perform it, or that every self-training system collapses. It makes a more specific claim about the route from recursive generative modelling to recursive self-improvement. An intelligence explosion requires an improvement operator. Closed-loop density matching does not supply one. Mechanism construction and non-circular evaluation are therefore not decorative additions to the Singularity story; they are part of what the positive-feedback premise still has to explain.

\section{Conclusions}

Recursive self-training should not be identified with recursive self-improvement. In the exact population limit, self-training under log-loss reproduces the current distribution. Under finite replacement, repeated resampling can reduce diversity and introduce stochastic distortion. External signals can stabilise the process, but their instantaneous proportion is not decisive. Cumulative corrective influence, sample size, retention and the form of the update determine the simple dynamics analysed here.

This correction does not weaken the architectural point. It sharpens it. The relevant danger is not the problem of vanishing real-data percentage, as a universal law but of epistemic closure: a loop in which candidate generation, training targets and evaluation become increasingly endogenous. A highly autonomous system need not be epistemically closed if it continues to interact with the world, use invariant verifiers or preserve other independent constraints.

The information-theoretic clarification is equally important. DPI and Levin non-growth remain fully valid. Their role is foundational: fixed processing cannot create information about an external mechanism from nothing. Their methodological force is limited in the finite empirical regime because they do not tell us which representation, program search or computational budget will expose useful structure already present in partial observations. Resource-bounded accessibility can increase while the global information ceiling remains unchanged.

CTM, BDM and AID should therefore be understood as tools for mechanism search and structural inference where others can exist, but not as exceptions to information conservation and not as substitutes for empirical grounding. Their potential contribution is to organise a search over executable generative explanations and to make structural hypotheses available for perturbation and testing. They can fail, and their outputs require independent validation.

A further conclusion follows for the relation between LLMs and Solomonoff induction. Cross-entropy, negative log-likelihood, maximum likelihood and direct equivalents such as perplexity-based model comparison optimise predictive fit to token distributions; they do not, on their own, supply the defining Solomonoff mixture over generative programs weighted by algorithmic description length. The fact that an LLM chooses its next token from a learned conditional distribution therefore does not make it a Solomonoff predictor. A sufficiently capable neural system may learn a bounded algorithmic procedure internally, but that would be an additional learned capability, not a property guaranteed by the likelihood objective or the softmax next-token interface. Section~\ref{sec:servedllm} shows that on the systems now advanced as approximate Solomonoff predictors this additional capability is, at present, not in evidence where it can be measured: across twenty-six served checkpoints none holds the dominance constant of a mixture that carries a complexity-blind code at weight $0.10$, the checkpoints nearest it settling to a constant two to four times larger and the others to none over the horizon read; every checkpoint puts mass off the declared class; and no effect of scale or post-training on allocation is established. The argument of this paper is therefore not only that likelihood training fails to \emph{entail} Solomonoff induction, but that none of the models in question exhibits its guarantee at the constant of the reference mixture, and several do not exhibit it at all.

This conclusion is compatible with the constructive result of Wan and Mei when their computable surrogate is distinguished from the likelihood objective itself~\cite{wanmei2025,wanmei2026}. Their construction shows how a trained LLM, arithmetic coding, prefix encoding and universal-machine machinery can jointly define a computable program family whose mass is contained within the Solomonoff mixture. That is an important route from an ordinary predictor towards a Solomonoff-style surrogate. The present result identifies where the extra inductive structure enters. Cross-entropy determines the fit of the model to the supplied distribution; the program-based coding construction determines how that fitted model contributes algorithmic mass. There is therefore no contradiction at this level. The present paper refines the claim by separating likelihood optimisation from the additional conditions under which the resulting computable surrogate can be treated as close to the full Solomonoff prior. Only the stronger assertion that likelihood minimisation alone guarantees convergence to the universal semimeasure would go beyond what the cross-entropy identity establishes.

This also identifies where much of the inductive burden resides. Likelihood-based learning can acquire powerful structural biases through architecture, representation, data selection, weighting, curricula and other transformations of the effective training distribution, but those interventions should be counted as part of the learning system rather than attributed to the likelihood-based objective itself. If an external process must continually reshape the probability mass seen by the learner in order to obtain the desired inference, then part of the intelligence of the resulting system resides in that curation process. Solomonoff induction relocates more of this burden towards computation over generative hypotheses. The contrast is not that one paradigm uses data and the other computation, but that they distribute the inferential burden differently between observations, external curation, architecture and internal mechanism search.

This difference also changes the significance of scaling computational power. More compute applied to a fixed likelihood objective can reduce optimisation error, support larger parameterisations or enable additional procedures, but the objective does not thereby become Solomonoff induction. In an open-ended CTM-based approximation, by contrast, additional computational resources can be used directly to enlarge the program and runtime classes contributing algorithmic mass. BDM and AID can extend this search towards larger structures and their perturbation. The resulting positive-feedback possibility is therefore not simply that a larger predictor fits more data, but that increased capability may enlarge the mechanism-search process that constructs candidate explanations for subsequent independent testing.

The algorithmic route considered here is constructive rather than prohibitive. CTM provides the direct connection to finite approximation of algorithmic probability; an open-ended CTM process can expand the program  classes and runtime budgets it explores. BDM can extend the analysis to larger structured objects, and AID can support perturbational analysis and reprogram mability of candidate mechanisms. The resulting architecture should not be identified with exact Solomonoff induction, which remains uncomputable. It is better characterised as a progressively extensible, resource-bounded Solomonoff-style estimator coupled to structural analysis. When its candidate mechanisms are tested against independent evidence or formal constraints, such an architecture supplies a concrete way in which more computational capability can enlarge mechanism search and potentially improve the process that generates further improvements.

Recursive LLM generation provides an extraordinary engine for producing candidates, but candidate production is only one part of that loop. Sustained self-improvement also requires a way to determine which changes constitute real improvement, to derive consequences that reach beyond self-consistency, and to test them against evidence or invariant criteria.

In that sense, the route from present recursive distributional learning to an intelligence explosion remains incomplete without a functional layer of symbolic model synthesis. That layer need not look like classical symbolic AI and need not be separate from a neural system. What matters is the operation: construct mechanisms, derive consequences, test them non-circularly, and retain what survives. Without that operation, recursion alone is not self-improvement.

Our results contradict some recent literature on `predictive' statistical compression aiming to show that LLMs are Solomonoff induction estimators. As we argue, such type of compression not only is difficult to compare being lossy but also does not and cannot on its own establish algorithmic comprehension. Del\'etang et al.~\cite{deletang2024compression} relate likelihood to coding efficiency; the stronger interpretation advanced in Li et al.~\cite{li2025understanding} does not follow from compression gains alone. We believe that the claim to surpass Shannon's bound is likewise unsupported: outperforming particular statistical compressors does not violate the source-coding theorem~\cite{coverthomas2006}. Our own work and ideas on the relationship between compression and comprehension~\cite{zenil2020comprehension} frames understanding through compact generative explanations, while our \emph{SuperARC}~\cite{hernandez2026superarc}  test, evaluates model abstraction and recursive planning (prediction by simulation). The present work supplies a complementary theoretical and empirical distinction: likelihood optimisation alone does not entail Solomonoff induction, pure self-imitation supplies no intrinsic improvement operator, and our experiments here and in ~\cite{hernandez2026superarc} separate improved predictive coding from improved extrapolation. Together, these results challenge the identification of coding gains with algorithmic understanding~\cite{deletang2024compression,li2025understanding}.

\section*{Use of AI tools disclosure statement}

Generative AI tools were used to assist with language editing, LaTeX formatting, checking derivations in the Sup Mat, proposing objections to arguments, and organising the material. All mathematical arguments, proofs, results, and interpretations originated and were verified by the authors, who assumes full responsibility for the manuscript.

\newpage

\appendix
\section{Supplementary Material}

From \eqref{eq:mix}, subtracting $P$ isolates the error relative to the reference distribution:
\[
Q_{t+1}-P=(1-\alpha_t)(Q_t-P).
\]
A single grounded update therefore multiplies the discrepancy by $1-\alpha_t$, so every positive grounding weight contracts it at that step. Repeated substitution gives \eqref{eq:beta}. If some $\alpha_t=1$, the residual becomes zero immediately. Otherwise the standard infinite-product criterion gives $\prod_t(1-\alpha_t)=0$ exactly when $\sum_t -\log(1-\alpha_t)=\infty$. Since $-\log(1-x)\geq x$ for $0\leq x<1$, divergence of $\sum_t\alpha_t$ is sufficient. If $\sum_t\alpha_t<\infty$, then eventually $\alpha_t$ is small and $-\log(1-\alpha_t)\leq 2\alpha_t$, so the logarithmic series converges and the product remains positive. Passing to logarithms is what converts the cumulative product into a sum and separates schedules whose many small corrections accumulate without bound from those whose total corrective influence remains finite.

Let $X\sim\mathrm{Multinomial}(N,q)$ and $q'=X/N$. For each coordinate,
\[
\E[(q_i')^2]=q_i^2+\frac{q_i(1-q_i)}{N}.
\]
The first term is the squared mean of the empirical frequency $q_i'$, while $q_i(1-q_i)/N$ is its sampling variance. Their sum gives the second moment needed to track diversity under resampling. Summing over $i$ and using $\sum_iq_i=1$ yields
\[
\E[V(q')\mid q]=\left(1-\frac1N\right)V(q).
\]
Thus finite sampling reduces expected Gini--Simpson diversity by the factor $1-1/N$ in one replacement step. Iteration gives
\[
\E V(q_t)=V(q_0)\prod_{s<t}\left(1-\frac1{N_s}\right).
\]
The product records the cumulative diversity retained after applying the same one-step contraction through all generations up to time $t$. Each coordinate is also a bounded martingale because $\E[q_{t+1,i}\mid q_t]=q_{t,i}$. Hence the coordinates converge almost surely. If $\sum_t1/N_t=\infty$, the product above tends to zero, and bounded convergence implies that the limiting diversity is zero almost surely. On the finite simplex this means convergence to a vertex. The limiting vertex is random rather than selected by an improvement criterion. Equivalently, resampling introduces fluctuations without systematically pushing any coordinate upwards or downwards; convergence to a vertex means eventual fixation on a state in which one outcome carries all probability mass, not directed improvement.

For grounded multinomial replacement, let
\[
r_t=\alpha_t p+(1-\alpha_t)q_t,
\qquad
q_{t+1}=X_t/N_t,
\qquad
X_t\sim\mathrm{Multinomial}(N_t,r_t).
\]
Thus $r_t$ is the mixture actually sampled at step $t$, and $q_{t+1}$ is the empirical frequency vector obtained from $N_t$ multinomial draws from it. Writing $e_t=q_t-p$ gives
\[
\E[e_{t+1}\mid q_t]=(1-\alpha_t)e_t,
\]
so the conditional mean error relative to the reference is contracted by the factor $1-\alpha_t$. The corresponding squared-error recursion is
\[
\E[\|e_{t+1}\|_2^2\mid q_t]
=(1-\alpha_t)^2\|e_t\|_2^2
+\frac{1-\|r_t\|_2^2}{N_t}.
\]
It separates deterministic contraction from finite-sampling noise: the first term is the squared error remaining after grounding, while the second is the new variance introduced by estimating $q_{t+1}$ from only $N_t$ samples. Let $x_t=\E\|e_t\|_2^2$ and $\gamma_t=2\alpha_t-\alpha_t^2$. Since $1-\|r_t\|_2^2\leq1$,
\[
x_{t+1}\leq(1-\gamma_t)x_t+\frac1{N_t}.
\]
The upper bound reduces the vector recursion to a scalar one: the previous expected squared error is contracted by $1-\gamma_t$, while $1/N_t$ bounds the fresh sampling noise. If $\sum_t\alpha_t=\infty$ and $N_t\alpha_t$ grows without bound, then $\sum_t\gamma_t=\infty$ and $(1/N_t)/\gamma_t$ tends to zero. The two limits have complementary meanings: contraction keeps accumulating while finite-sampling noise becomes negligible relative to it. Given any $\varepsilon>0$, eventually $1/N_t\leq\varepsilon\gamma_t$, and hence
\[
x_{t+1}-\varepsilon\leq(1-\gamma_t)(x_t-\varepsilon).
\]
Once the noise is bounded by an arbitrarily small multiple of the contraction strength, subtracting $\varepsilon$ leaves another contracting recursion. Iteration and the vanishing product $\prod_t(1-\gamma_t)$ imply $\limsup_t x_t\leq\varepsilon$. Since $\varepsilon$ is arbitrary, the limiting upper bound is zero and $x_t$ tends to zero, establishing mean-square convergence. The schedule $\alpha_t=1/(t+2)$ and $N_t=(t+2)^2$ therefore supplies a direct finite-sample counterexample to the claim that a vanishing grounding fraction necessarily causes collapse.

\end{document}